\documentclass[10pt,twocolumn,reqno, aps,prl,superscriptaddress]{revtex4-2}%
\usepackage{amsmath}
\usepackage[pdftex]{graphicx}
\usepackage{float}
\usepackage{color}
\usepackage{rotating}
\usepackage[breaklinks=true,colorlinks,citecolor=blue,linkcolor=blue,urlcolor=blue]%
{hyperref}
\usepackage{graphicx}
\usepackage{natbib}
\usepackage{dcolumn}
\usepackage{bm}
\usepackage{hyperref}
\usepackage[caption=false]{subfig}
\usepackage{epstopdf}
\usepackage{ulem}
\usepackage{amsfonts}
\usepackage{amssymb}%
\providecommand{\U}[1]{\protect\rule{.1in}{.1in}}
\providecommand{\U}[1]{\protect\rule{.1in}{.1in}}
\begin{document}

\title{Theory of the Magnon Parametron}
\author{Mehrdad Elyasi}
\affiliation{Institute for Materials Research, Tohoku University, Sendai 980-8577, Japan}
\author{ Eiji Saitoh}
\affiliation{Institute for Materials Research, Tohoku University, Sendai 980-8577, Japan}
\affiliation{WPI Advanced Institute for Materials Research, Tohoku University, Sendai 980-8577, Japan}
\affiliation{Center for Spintronics Research Network, Tohoku University, Sendai 980-8577, Japan}
\affiliation{Department of Applied Physics, University of Tokyo, Hongo, Tokyo 113-8656, Japan}
\affiliation{Advanced Science Research Center, Japan Atomic Energy Agency, Tokai 319-1195, Japan}
\author{Gerrit E. W. Bauer}
\affiliation{Institute for Materials Research, Tohoku University, Sendai 980-8577, Japan}
\affiliation{WPI Advanced Institute for Materials Research, Tohoku University, Sendai 980-8577, Japan}
\affiliation{Center for Spintronics Research Network, Tohoku University, Sendai 980-8577, Japan}
\affiliation{Zernike Institute for Advanced Materials, University of Groningen, 9747 AG Groningen, Netherlands}

\date{\today }

\begin{abstract}
The `magnon parametron' is a ferromagnetic particle that is parametrically
excited by microwaves in a cavity. Above a certain threshold of the microwave
power, a bistable steady state emerges that forms an effective Ising spin. We
calculate the dynamics of the magnon parametron as a function of microwave
power, applied magnetic field and temperature for the interacting magnon
system, taking into account thermal and quantum fluctuations. We predict three
dynamical phases, viz. a stable Ising spin, telegraph noise of thermally
activated switching, and an intermediate regime that at lower temperatures is
quantum correlated with significant distillible magnon entanglement. These
three regimes of operation are attractive for alternative computing schemes.

\end{abstract}
\maketitle

An Ising spin is a magnetic moment with a large uniaxial anisotropy that
reduces the quantum degree of freedom of the Heisenberg spin on the Bloch
sphere to just two, i.e. up and down. More generally, the term is used for any
bistable system with a phase space of two distinct and stable configurations.
For example, the magnetization of a fixed ferromagnetic needle that can point
only into the two directions that minimizes the free energy is a (pseudo)
Ising spin. An Ising spin with noise-activated transitions can operate as a
probabilistic bit (p-bit), which in its steady state is a statistical mixture
of the two levels. Ising spins are not useful as qubits because the large
energy barrier prevents spin rotations on the Bloch sphere. Nevertheless,
interactions with other degrees of freedom can induce quantum coherence of the
Ising up and down spins and entanglement with other excitations. An ensemble
of spins in these three regimes form a platform for unconventional computing
algorithms. Switchable, but thermally stable, Ising spins are elements of
\textquotedblleft Ising machines\textquotedblright\ that can solve hard
optimization problems
\cite{Yamaoka2015,Inagaki2016,MacMahon2016,Inagaki2016_1,Pierangeli2019},
while network of p-bits can factorize large integers \cite{Borders2019}. A
relatively large ($\sim2000$) and highly connected network of pseudo Ising
spins with phase measurement and feedback was implemented by a train of
optical parametric oscillators \cite{Inagaki2016,MacMahon2016,Inagaki2016_1}.
However, optical implementations have a large footprint and are not scalable.
Quantum coherent networks are even more difficult to realize, but they can
perform additional tasks such as quantum annealing, adiabatic evolution, or
gated quantum operations
\cite{DiVincenzo1995,Nielsen2009,Farhi2001,Albash2018,Johnson2011,Boixo2014}.

Parametric pumping is a standard method to excite large oscillations in a
harmonic oscillator by a phase matched drive at twice the resonance frequency
$\omega_{0}$. When a harmonic oscillator with Hamiltonian $\hbar\omega
_{0}a^{\dag}a$, where $a^{\dag}\left(  a\right)  $ creates (annihilates) a
boson, has non-linear interaction with photons, it can be driven into an
instability by the parametric term $Pe^{2i\omega_{0}t}a^{\dag}a^{\dag
}+\mathrm{H.c.}$, when the classical amplitude $P$ exceeds a certain
threshold. In the steady state, the mean field $\langle a\rangle$
spontaneously acquires either one of the energetically equivalent phases of
$\phi_{p}/2+0$ or $\phi_{p}/2+\pi$, where $\phi_{p}=\mathrm{\arg}%
Pe^{2i\omega_{0}t}$ and $2\operatorname{mod}\mathrm{\arg}\left[  \langle
a\rangle,2\pi\right]  =\operatorname{mod}\left[  \phi_{p},2\pi\right]  $,
which can be mapped on the two states of a pseudo Ising spin. Such oscillators
can be realized by optical \cite{Inagaki2016,MacMahon2016,Inagaki2016_1},
electromechanical \cite{Mahboob2016}, or magnetic \cite{Makiuchi2021} systems.
Makiuchi et al. \cite{Makiuchi2021} demonstrated a \textquotedblleft magnon
parametron\textquotedblright\ on a disk of the magnetic insulator yttrium iron
garnet (YIG) that also showed the stochastic behavior expected for a p-bit.

A Hilbert space of a quantum system is `discrete' when its dimension is
countable, e.g., two for a spin-$1/2$ system. It is called `continuous' when
uncountably infinite, e.g. when spanned by position and momentum variables of
a harmonic oscillator. The lowest number states of the Kittel magnons, i.e.
the quanta of the uniform precession of the magnetic order
\cite{Huebl2013,Zhang2014,Tabuchi2014} enable `discrete variable' quantum
information processing \cite{Tabuchi2015,Lachance2020}. However, since their
anharmonicity is small, an auxiliary superconducting qubit is required to
manipulate the quantum states of the lowest magnon levels. On the other hand,
strongly driven magnons alone offer `continuous variable' quantum information
such as entanglement, photon squeezing, and antibunching
\cite{Li2018,Elyasi2020,Yuan2020}. Magnons are also set apart from e.g.
phonons \cite{Aspelmeyer2014} by their highly tunable, anisotropic and
non-monotonic dispersions. Parametric excitation of magnets generates large
magnon numbers that live long enough to form Bose-Einstein condensates
\cite{Demidov2007,Demidov2008,Serga2013,Bozhko2016}. We are especially
interested in the efficient and distillable entanglement of magnons
\cite{Elyasi2020} in the parametrically excited regime.

In this Letter, we address the theory of the `magnon parametron'
\cite{Makiuchi2021}, i.e. a thin film magnetic disc that is parametrically
excited in a microwave cavity. We show that it can operate as an Ising spin
that is tuned between deterministic, stochastic, and quantum regimes, which
should be considered seriously as a platform for alternative computing
technologies. The Suhl instability \cite{Suhl1957,Elyasi2020}, i.e. the decay
of the uniform (Kittel) magnon into a pair of magnons with opposite momenta
$\pm{k}\neq0$, can be reached by driving magnon parametrically by cavity-mode
microwaves with a small amplitude when quality factors are high. At a
classical fixed point of the magnetization dynamics and at cryogenic
temperatures we predict a distillable quantum entanglement \cite{Elyasi2020}.
The limit-cycle dynamics at slightly higher photon amplitudes enables the
observed stochastic switching between the Ising spin states
\cite{Makiuchi2021} only when the `wing' magnons are involved.

\begin{figure}[ptb]
\includegraphics[width=0.5\textwidth]{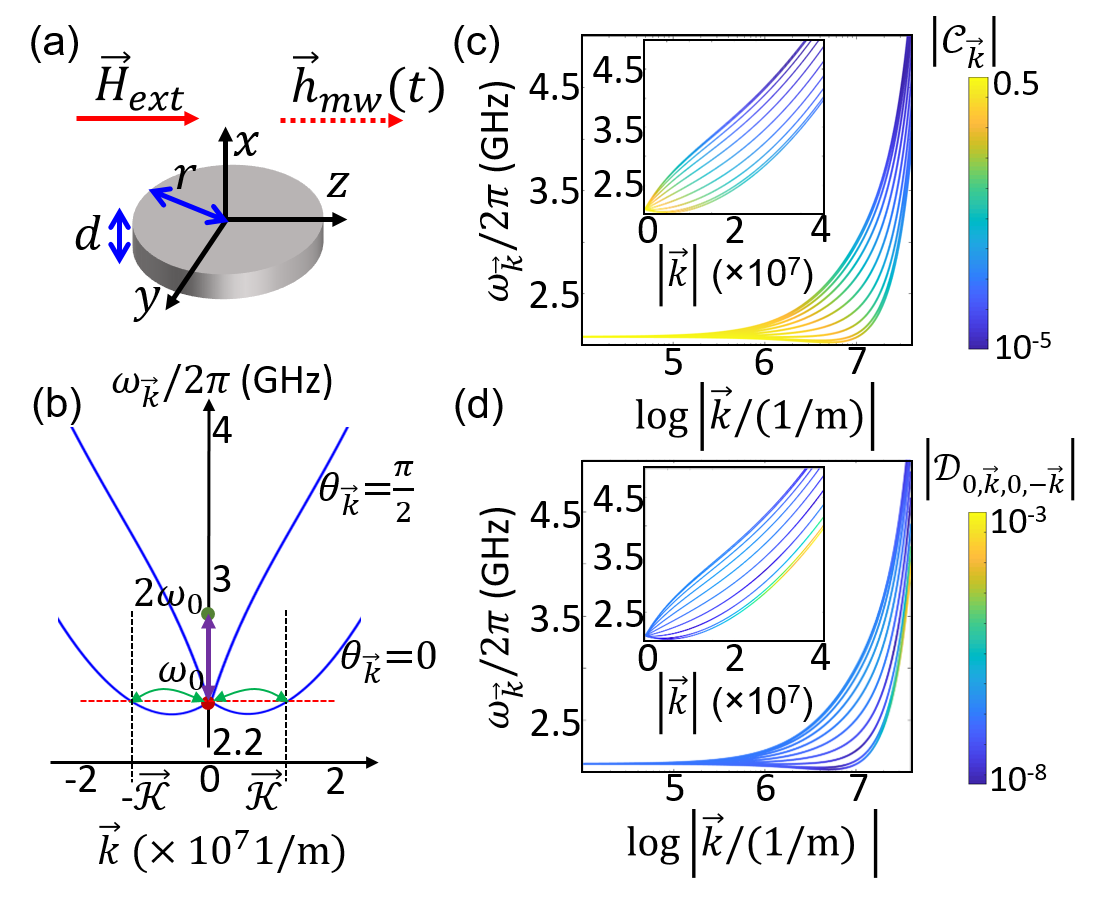}\caption{Model. (a) A
magnetic disk of thickness $d$ and radius $r$ under static and microwave
magnetic fields. (b) The dispersion envelope of magnons with constant
magnetization along $x$ for $d=50\,$nm and $r=100\,\mathrm{\mu}$m. The green
arrows indicate 4-magnon scattering processes involving the Kittel mode. The
purple line indicates parametric pumping of the Kittel mode. (c) The
parametric excitation coefficient amplitude $\mathcal{C}_{\vec{k}}$ of the
magnon pairs overlaid on the dispersion for several values of $\theta_{\vec
{k}}$ from $0$ to $\pi/2$, only for the modes nodeless along the thickness.
(d) Similar to (c) but for the 4-magnon scattering coefficient $\mathcal{D}%
_{0,\vec{k},0,-\vec{k}}$. The insets of (c) and (d) plot the data of the main
panels on a linear momentum scale..}%
\label{fig1}%
\end{figure}

\textit{Model -} Figure \ref{fig1}(a) sketches a thin ferromagnetic disk of
thickness $d$ and radius $r$, uniformly magnetized along the in-plane magnetic
field $\vec{H}_{ext}\Vert\hat{z}$. The microwave magnetic field $\vec{h}%
_{mw}\Vert\hat{z}$ of a cavity or a coplanar waveguide mode with frequency
$\omega_{p}$ is polarized \emph{along} the magnetization. Figure \ref{fig1}(b)
shows the magnon frequency dispersions $\omega_{\vec{k}}$ of a typical YIG
disk of $d=50\,$nm, corresponding to $\theta_{\vec{k}}=0$ ($\vec{k}\Vert
\hat{z}$) and $\theta_{\vec{k}}=\pi/2$ ($\vec{k}\perp\hat{z}$) for in-plane
wave vectors and constant magnetization along $\hat{x}$, i.e. the lowest
magnon subband \cite{Kalinikos1986,Hurben1995,Rezende2009}. $\omega_{0}$ is
the frequency of the Kittel mode. A node along $\hat{x}$ blue-shifts the
entire dispersion $\sim4\gamma D/d^{2}\sim0.2\,$GHz relative to the bulk value
of Fig. \ref{fig1}(b), where $\gamma=26\,\text{GHz}$/$\text{T}$ is the
gyromagnetic ratio and $D=2\times10^{-17}\,\text{T}\text{m}^{2}$ is the
exchange stiffness \cite{Stancil2009}. We restrict our study to the lowest
subband since for the chosen dimensions $\omega_{\vec{k}_{\mathrm{dip}}%
}+4\gamma D/d^{2}>\omega_{0}$, where $\omega_{\vec{k}_{_{\mathrm{dip}}}}$ is
the frequency minimum caused by the magnetodipolar interaction. The two
valleys in the magnon dispersion are essential for the Suhl instability and
exist when $r\gtrapprox0.5\,\mathrm{\mu}$m.

Figure \ref{fig1}(b) sketches the Kittel mode and a degenerate pair of magnons
with wave vector $\pm\mathcal{K}$ as well as the four-magnon scattering
process relevant to a Suhl instability. $\vec{h}_{mw}(t)$ with frequency
$\omega_{p}=2\omega_{0}$ parametrically interacts with the Kittel mode and the
degenerate magnon pairs. Our Hamiltonian contains the leading terms of the
Holstein-Primakoff expansion of the Heisenberg model, including all 4-magnon
interactions \cite{Krivosik2010,Elyasi2020,Rezende2020}
\begin{align}
H  &  =H_{m,L}+H_{m,NL}+H_{mp}+{H_{p}},\nonumber\\
H_{m,L}  &  =\sum_{\vec{k}}\omega_{\vec{k}}c_{\vec{k}}^{\dag}c_{\vec{k}%
}\nonumber\\
H_{m,NL}  &  =\sum_{\vec{k}}\left\{  \mathcal{D}_{\vec{k},\vec{k},\vec{k}%
,\vec{k}}c_{\vec{k}}^{\dag}c_{\vec{k}}c_{\vec{k}}^{\dag}c_{\vec{k}}+\right.
\nonumber\\
&  \sum_{\vec{k}^{\prime\prime}}\left[  \left(  1-\delta_{\vec{k},\vec
{k}^{\prime\prime}}\right)  \mathcal{D}_{\vec{k},\vec{k},\vec{k}^{\prime
\prime},\vec{k}^{\prime\prime}}c_{\vec{k}}^{\dag}c_{\vec{k}}c_{\vec{k}%
^{\prime\prime}}^{\dag}c_{\vec{k}^{\prime\prime}}+\right. \nonumber\\
&  \left.  \left.  \left(  1-\delta_{|\vec{k}|,|\vec{k}^{\prime\prime}%
|}\right)  \frac{1}{2}\mathcal{D}_{\vec{k},\vec{k}^{\prime\prime},-\vec
{k},-\vec{k}^{\prime\prime}}c_{\vec{k}}^{\dag}c_{-\vec{k}}^{\dag}c_{\vec
{k}^{\prime\prime}}c_{-\vec{k}^{\prime\prime}}\right]  \right\}  ,\nonumber\\
H_{mp}  &  =\sum_{\vec{k}}\frac{1}{2}\left(  1+\delta_{\vec{k},0}\right)
\left(  \mathcal{G}_{\vec{k}}bc_{-\vec{k}}^{\dag}c_{\vec{k}}^{\dag
}+H.c.\right)  ,\nonumber\\
H_{p}  &  =\omega_{p}b^{\dag}b+E\left(  b^{\dag}+b\right)  , \label{eq1}%
\end{align}
where
$\mathcal{G}_{\vec{k}}=\gamma\mathcal{C}_{\vec{k}}\sqrt{\hbar\mu_{0}\omega
_{p}/2V_{p}}$, $b$ is the photon annihilation operator, $V_{p}$ is the cavity
mode volume,
$m_{z}=1-\sum_{\vec{k}}\left[  c_{\vec{k}}^{\dag}c_{\vec{k}}+\left(
\mathcal{C}_{\vec{k}}c_{-\vec{k}}^{\dag}c_{\vec{k}}^{\dag}+\mathrm{H.c.}%
\right)  \right]  /S$ is the $\hat{z}$ component of the magnetization unit
vector, the total spin $S=M_{s}V_{m}/2\pi\gamma\hbar$, $V_{m}$ is the volume
of the sample, $M_{s}$ is the saturation magnetization. The coefficients
$\mathcal{C}$ and $\mathcal{D}$ are complicated but well known
\cite{Rezende2020,Krivosik2010}.

\begin{figure}[t]
\includegraphics[width=0.5\textwidth]{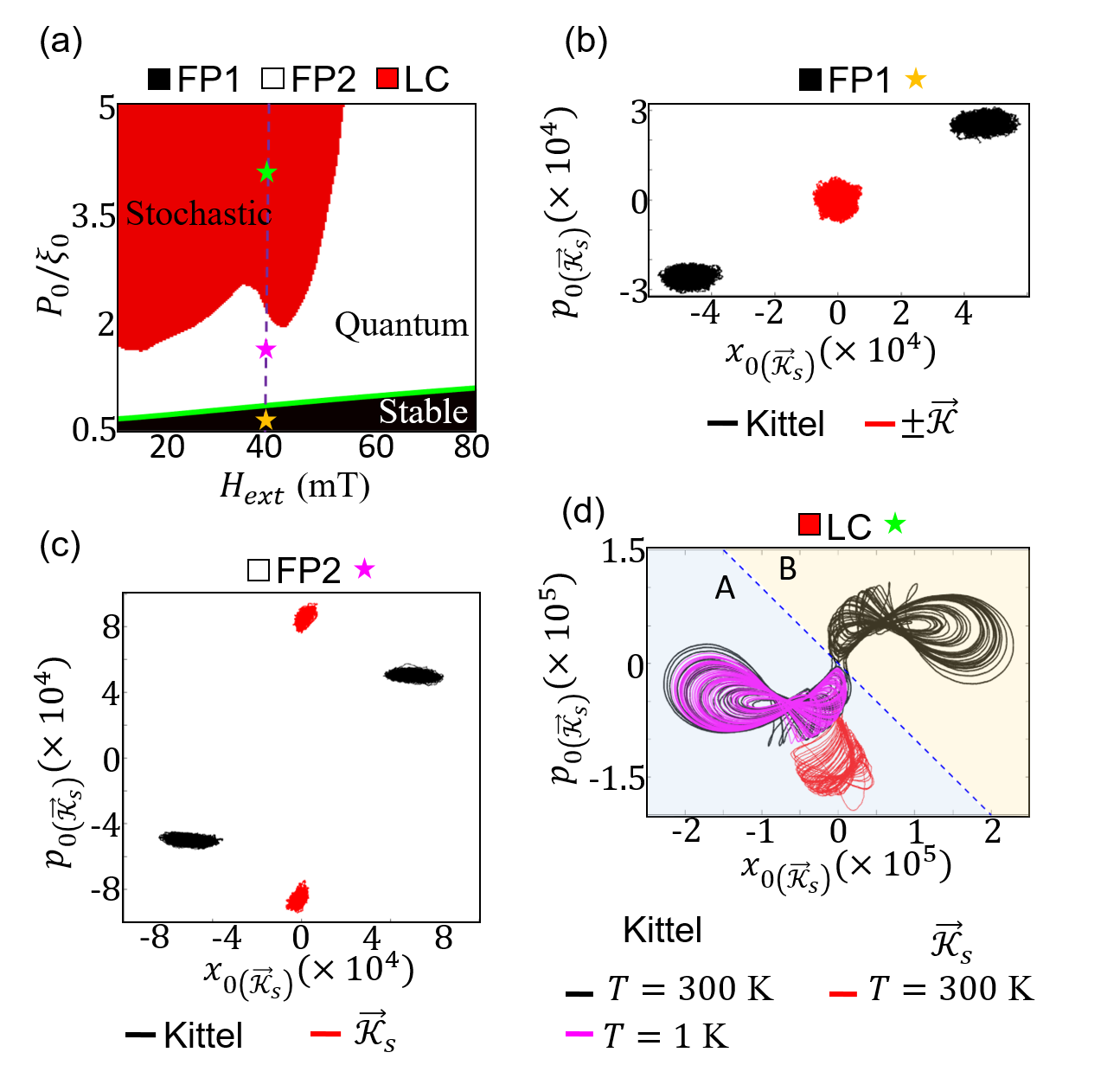}\caption{Calculated
steady-state dynamics of a magnet in a microwave cavity. (a) The dependence of
steady-state class on the Kittel mode amplitude $P_{0}$ driven by the
mircrowaves and the dc magnetic field $H_{\mathrm{ext}}$, labeled as `Stable',
`Quantum', and `Stochastic'. (b)-(d) Examples for the three distinct classes,
corresponding to the stars of the same color in in (a), $P_{0}/\xi
_{0}=0.7,\,1.7,\,3.8$, respectively, while $H_{ext}=40\,$mT. (b) FP1: Fixed
point, the Kittel mode parametrically driven beyond threshold, while the
$\vec{\mathcal{K}}_{s}$ pair at vacuum. FP2: Fixed point, the Kittel mode and
$\vec{\mathcal{K}}_{s}$ standing wave parametrically and Suhl instability
driven, respetively. In (b) and (c), $T=3\times10^{5}\,$K for clarity. (d) LC:
Limit cycle to chaos. A case with large transition rate from one attractor
region of the Kittel mode to the other, at $T=300\,$K, and no transition for
$T=1\,$K also shown. The two attractor regions A and B indicated. In (b) and
(c), green trajectories are for $\mathcal{K}$ standing wave. }%
\label{fig2}%
\end{figure}

\begin{figure}[t]
\includegraphics[width=0.5\textwidth]{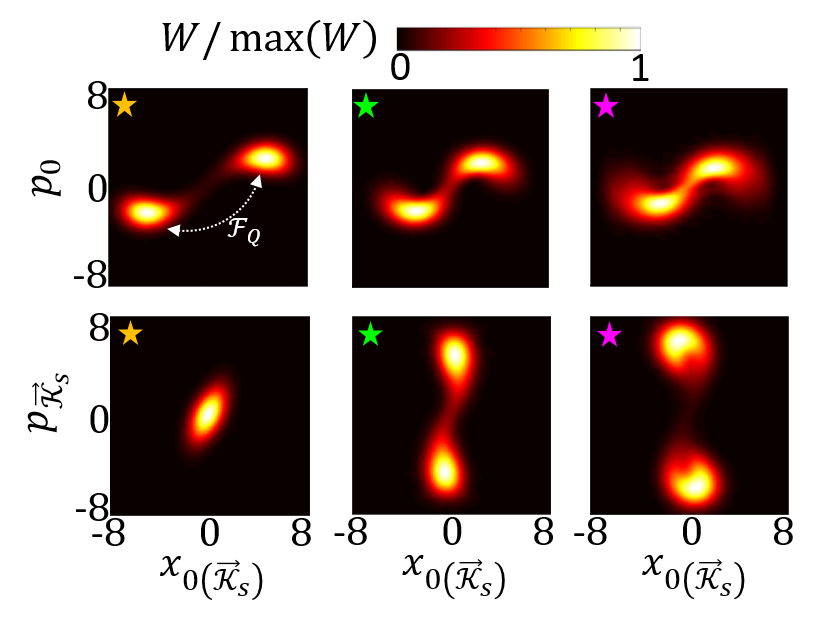}\caption{Quantum steady
states. The top (bottom) panels are the Wigner function of the Kittel mode
($\mathcal{K}$ standing wave) normalized by the maximum value. The left,
middle, and right panels correspond to the star of the same color as in Fig.
\ref{fig2}(a), as well as Figs. \ref{fig2}(b)-(d), i.e., in the FP1 (Ising),
`Quantum' (FP2), and `Stochastic' (LC) regions, respectively. The scaling of
four-magnon scattering coefficients $\mathcal{Q}=5\times10^{8},\,2\times
10^{9},\,4\times10^{9}$, from left to right panels, respectively.}%
\label{fig3}%
\end{figure}

Here we pump the precession cone angle of spin waves by the microwave magnetic
field along the magnetization. The photon drive $E$ with $\omega_{p}%
\sim2\omega_{0}$ leads to a coherent photon field $\langle b\rangle=\beta$
such that $\mathcal{G}_{\vec{k}}\beta c_{-\vec{k}}^{\dag}c_{\vec{k}}^{\dag
}+\mathrm{H.c}.$ can parametrically excite the Kittel mode and magnon pairs
with $\omega_{\vec{k}}\sim\omega_{p}/2$. When increasing $E,$ the mode with
the largest $\mathcal{G}_{\vec{k}}\propto\mathcal{C}_{\vec{k}}$ becomes
instable at a critical value $\mathcal{G}_{\vec{k}}\beta=\xi_{\vec{k}}/2$,
where $\xi_{\vec{k}}\approx\alpha_{G}\omega_{\vec{k}}$ is the magnon
dissipation rate in terms of $\alpha_{G}$, the Gilbert damping constant.
$|\mathcal{C}_{\vec{k}}|$ in Figure \ref{fig1}(c) is maximal for small wave
vectors, which implies that the Kittel mode becomes instable first. The
\textquotedblleft drift\textquotedblright\ matrix $\mathcal{O}=\left[
i\omega_{0}-\xi_{0}/2,i\mathcal{G}_{0}\beta;-i\mathcal{G}_{0}\beta
,-i\omega_{0}-\xi_{0}/2\right]  $ of the linear equation of motion $\left[
\dot{c}_{0},\dot{c}_{0}^{\dag}\right]  ^{T}=\mathcal{O}\left[  {c}_{0},{c}%
_{0}^{\dag}\right]  ^{T}$ then acquires an eigenvalue with positive real part.
The so-called self-Kerr coefficient $\mathcal{D}_{0,0,0,0}$ governs steady
state Kittel magnon amplitude just above the threshold. Here we use rather
large damping parameter $\xi_{0}=5\,$MHz corresponding to $\alpha_{G}%
\sim2\times10^{-3}$ for computational convenience.

The parametrically driven Kittel mode excites other magnons via the
four-magnon scattering term $c_{0}^{\dag}c_{0}^{\dag}c_{\vec{k}}c_{-\vec{k}%
}+\mathrm{H.c.}$, where Fig. \ref{fig1}(d) plots the corresponding
coefficients $\left\vert \mathcal{D}_{0,\vec{k},0,-\vec{k}}\right\vert $. This
is another threshold process introduced first by Suhl \cite{Suhl1957}.
Ignoring the terms $c_{0}^{\dag}c_{0}c_{\pm\vec{k}}^{\dag}c_{\pm\vec{k}}$ for
the moment, the instability is reached when the amplitude of the Kittel mode
mean field $\left\vert \alpha_{0}\right\vert =\langle c_{0}\rangle=\left(
\sqrt{\xi_{\vec{k}}^{2}/4+\Delta_{\vec{k}}^{2}}/|\mathcal{D}_{0,\vec
{k},0,-\vec{k}}|\right)  ^{1/2}$. This happens first for the degenerate modes
with largest $\left\vert \mathcal{D}_{0,\vec{k},0,-\vec{k}}\right\vert $, i.e.
for $\theta_{\vec{k}}=0$ and large $\left\vert \vec{k}\right\vert $ and
thereby limits the Hilbert space to three modes, the parametrically pumped
Kittel mode and a pair of magnons with large wave vector $\pm\mathcal{\vec{K}%
}$. In the rotating frame of $\mathcal{\omega}_{p}/2,$ the Hamiltonian
(\ref{eq1}) reduces to $H^{\prime}=H_{m,L}^{\prime}+H_{m,NL}^{\prime}%
+P_{0}\left(  c_{0}^{\dag}c_{0}^{\dag}+\mathrm{H.c}.\right)  $, where
$H_{m,L}=\sum_{\vec{k}\in{\{0,\pm\mathcal{K}\}}}\Delta\omega_{\vec{k}}%
c_{\vec{k}}^{\dag}c_{\vec{k}}$, $\Delta\omega_{\vec{k}}=\omega_{\vec{k}%
}-\omega_{p}/2$, $H_{m,NL}^{\prime}=\left.  H_{m,NL}\right\vert _{\vec{k}%
\in\{0,\pm\mathcal{K}\}}$. The critical parametric excitation amplitude
$P_{0}=\xi_{0}/2$ corresponds to a photon amplitude $E\approx P_{0}\xi
_{p}/2\mathcal{G}_{0}$ and power $\mathcal{P}=E^{2}\hbar\omega_{p}/\xi
_{p}=P_{0}^{2}\xi_{p}\hbar\omega_{p}/4\mathcal{G}_{\vec{k}}^{2}$. Assuming
$\xi_{p}=10\,$kHz (photon quality factor $\sim10^{5}$), and $\mathcal{G}%
_{\vec{k}}/\mathcal{C}_{\vec{k}}=10$ which corresponds to a photon mode volume
$V_{p}\sim10^{-2}\,\mathrm{mm}^{3}$, for $P_{0}=\xi_{0}/2$, $\mathcal{P}%
\approx2\,\mathrm{\mu W}$ and for $P_{0}=5\xi_{0}$ (maximum value used in our
calculations), $\mathcal{P}\approx0.2\,$mW.

The (Lindblad) equation of motion of the density matrix $\rho$ with elements
$\rho_{i,j}=\left\vert i\right\rangle \left\langle j\right\vert $, where
$|i(j)\rangle$ is a many-body number (Fock) state of the magnon system, reads%
\begin{equation}
\dot{\rho}=-i\left[  H^{\prime},\rho\right]  +L_{d}, \label{eq2}%
\end{equation}
where
\begin{align}
L_{d}  &  =\sum_{\vec{k}\in\{0,\pm\mathcal{K}\}}\xi_{\vec{k}}\left[
n_{th}(\omega_{\vec{k}})\left(  c_{\vec{k}}\rho c_{\vec{k}}^{\dag}+c_{\vec{k}%
}^{\dag}\rho c_{\vec{k}}-\rho c_{\vec{k}}c_{\vec{k}}^{\dag}-\right. \right.
\nonumber\\
&  \left. \left. c_{\vec{k}}^{\dag}c_{\vec{k}}\rho\right)  + \frac{1}%
{2}\left(  2c_{\vec{k}}\rho c_{\vec{k}}^{\dag}-c_{\vec{k}}^{\dag}c_{\vec{k}%
}\rho-\rho c_{\vec{k}}^{\dag}c_{\vec{k}}\right)  \right]  , \label{eq3}%
\end{align}
is the dissipation operator of the magnons in contact with a thermal bath.
Here $n_{th}(\omega_{\vec{k}})=\left(  e^{\hbar\omega_{\vec{k}}/k_{B}%
T}-1\right)  ^{-1}$, $k_{B}$ is the Boltzmann constant, and $T$ is the bath
temperature. We disregard nonlinear radiative damping terms since
$|\gamma\mathcal{C}_{\vec{k}}\sqrt{\hbar\mu_{0}\omega_{p}/2V_{p}}|^{2}/\xi
_{p}\xi_{0}\ll1$ \cite{Kinsler1991}. Without drive, $\rho$ describes a magnon
gas\ at thermal equilibrium with the bath.

Next, we show our results for the driven steady state, quantify stochasticity,
and discuss quantum entanglement in our magnetic dot.

\begin{figure*}[t]
\includegraphics[width=1\textwidth]{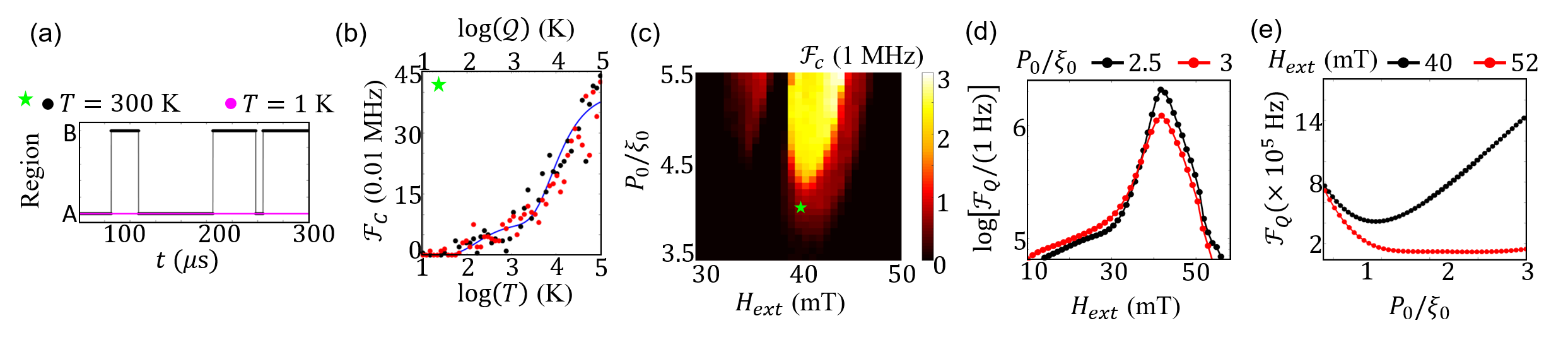}\caption{Stochasticity.
(a)-(c) From semi-classical, (d)-(f) from quantum calculations. (a) The region
of the Kittel mode state shown in Fig. \ref{fig2}(d) for $T=1\,$K and
$T=300\,$K. (b) The dependence of transition frequency $\mathcal{F}_{C}$ on
$T$ (black dots) and $\mathcal{Q}$ (red dots), for $\ H_{\mathrm{ext}%
}=40\,\text{mT},P_{0}/\xi_{0}=3.85$, as in (a) and Fig. \ref{fig2}(d). (c)
$\mathcal{F}_{C}$ as a function of $H_{\mathrm{ext}}$ and $P_{0}/\xi_{0}$, at
$T=3\times10^{5}\,$K. The green star is the same $(P_{0},H_{\mathrm{ext}})$
point as in the phase diagram plotted in Figs. \ref{fig2}(a) and (d). (d) The
dependence of tunneling frequency $\mathcal{F_{Q}}$ on $H_{ext}$ for two
values of $P_{0}/\xi_{0}=2.5,\,3$. (e) The dependence of $\mathcal{F}_{Q}$ on
$P_{0}/\xi_{0}$ for $H_{ext}=40,\,52\,$mT, respectively. The scaling
coefficient $\mathcal{Q}=5\times10^{9}$ in (d)-(e), and $T=0\,$K. }%
\label{fig4}%
\end{figure*}

\textit{Steady state classes -} We classify the steady state dynamics in terms
of a \textquotedblleft phase diagram\textquotedblright\ of our three-mode
system by the solutions of the Langevin equation of motion. Disregarding the
third and fourth order derivatives of the Wigner distribution functions as
described in the supplementary material (SM), Sec. I
\cite{SM,Carmichael1999,Walls2008} for small nonlinearities, simplifies the
equation of motion to $\dot{v}=-i[H,v]+\Gamma$, where $v=[x_{0},p_{0}%
,x_{\mathcal{K}},p_{\mathcal{K}},x_{-\mathcal{K}},p_{-\mathcal{K}}]$,
$x_{0(\pm\vec{\mathcal{K}})}=(c_{0(\pm\vec{\mathcal{K}})}+c_{0(\pm
\vec{\mathcal{K}})}^{\dag})/2$, $p_{0(\pm\vec{\mathcal{K}})}=-i(c_{0(\pm
\vec{\mathcal{K}})}-c_{0(\pm\vec{\mathcal{K}})}^{\dag})/2$, and $\Gamma$
represents fluctuating fields with Gaussian quantum statistics. We solve this
6-dimensional Langevin differential equation in real time starting from
appropriate initial conditions until the steady state is reached.

The microwaves parametrically excite the Kittel mode with detuning
$\Delta\omega_{0}=0$, and an amplitude $P_{0}$. The other control parameter is
the applied static magnetic field $H_{ext}$. The smallest positive solution
for $x=|\alpha_{0}|^{2}$ governs $\theta_{\vec{\mathcal{K}}}$ and
$|\vec{\mathcal{K}}|$ of the magnon pair that reaches the Suhl instability
first
\begin{gather}
\left(  \mathcal{D}_{0,\vec{\mathcal{K}},0,-\vec{\mathcal{K}}}^{2}%
-\mathcal{D}_{0,0,\vec{\mathcal{K}},\vec{\mathcal{K}}}^{2}\right)
x^{2}-2\Delta_{\vec{\mathcal{K}}}\mathcal{D}_{0,0,\vec{\mathcal{K}}%
,\vec{\mathcal{K}}}x\nonumber\\
-\xi_{\vec{\mathcal{K}}}^{2}/4-\Delta_{\vec{\mathcal{K}}}^{2}=0. \label{eq4}%
\end{gather}
Below the Suhl but above the parametric instability threshold $|\alpha
_{0}|^{2}=\sqrt{P_{0}^{2}-\xi_{0}^{2}/4}/2|\mathcal{D}_{0,0,0,0}|$.

With notation $c_{\pm\mathcal{K}}=|c_{\pm\mathcal{K}}|e^{i\phi_{\pm
\vec{\mathcal{K}}}},$ the four magnon scatterings fix the sum of the phases
$\phi_{+}=\phi_{\vec{\mathcal{K}}}+\phi_{-\vec{\mathcal{K}}}$, but the
difference $\phi_{-}=\phi_{\vec{\mathcal{K}}}-\phi_{-\vec{\mathcal{K}}}$ is
not uniquely determined \cite{Zakharov1974}. The magnetic disc has a large but
finite radius, that strictly speaking splits the continuum of state by
$\triangle\nu\sim2n\gamma D/r^{2}\sim10^{4}\text{Hz}$, where $n=\lfloor
2r\mathcal{K}\rfloor\sim4000$. Since $\triangle\nu\ll\xi_{0},$ the spectrum is
still quasi-continuous, but the Kittel mode decays not into two propagating,
but a single standing wave mode. This can be formalized by combining the
$\pm\vec{\mathcal{K}}$ pair of propagating waves as $c_{\pm\vec{\mathcal{K}}%
}=c_{\vec{\mathcal{K}}_{s}}e^{\mp iq/2}$ \cite{Bryant1988,Zakharov1974}, where the phase
$q=\phi_{-}$ is a free phase that governs the position of the standing wave
nodes and $\vec{\mathcal{K}}_{s}$ is a standing wave index. This reduction of
a three-partite into a two-partite problem simplifies the quantum regime calculations.

Figure \ref{fig2}(a) shows the steady state classes as a function of
$H_{\mathrm{ext}}$ and $P_{0}$, obtained numerically for $T=0\,$K. The green
line in Fig. \ref{fig2}(a) is an analytic solution of Eq. (\ref{eq4}) using
the four-magnon scattering parameters of the unstable mode for each
$H_{\mathrm{ext}}$. The phase-space dynamics of each class are illustrated by
Figs. \ref{fig2}(b)-(d) for a fixed magnetic field. Figures \ref{fig2}(b) and
(c) show trajectories in the time interval $t=50-80\,\mathrm{\mu}$s, starting
from 100 random initial values of $\phi_{0(\vec{\mathcal{K}}_{s})}$ in
$c_{0(\vec{\mathcal{K}}_{s})}=e^{i\phi_{0(\vec{\mathcal{K}}_{s})}}$ at
$t=0\ $and a high temperature $T=3\times10^{5}\,$K to emphasize the dynamic
stability. The trajectories are depicted in $(x_{0(\vec{\mathcal{K}}_{s}%
)},p_{0(\vec{\mathcal{K}}_{s})})$ phase space. We observe two distinct classes
that are characterized by Kittel mode fixed-points FP1 and FP2. For a given
$H_{\mathrm{ext}}$ and small $P_{0}>\xi_{0}/2$ (FP1) the Kittel mode has two
equivalent stable fixed points (Ising spin up and down), while $\vec
{\mathcal{K}}_{s}$ standing wave mode only fluctuates around the origin [see
Fig. \ref{fig2}(b)]. When $P_{0}$ satisfies Eq. (\ref{eq4}), the Suhl
instability drives the $\pm\mathcal{K}$ pair, leading to the FP2 steady state
in which the Kittel fixed points persist, and $\vec{\mathcal{K}}_{s}$ settles
at a fixed point away from the origin with a phase spontaneously chosen out of
two mirror symmetric values [see Fig. \ref{fig2}(c)]. We note that even though
FP2 is labeled \textquotedblleft quantum\textquotedblright, at the chosen high
temperatures all quantum correlations are of course washed out (see below). A
third distinct class without a stable fixed point is the limit cycle (LC)
illustrated in Figure \ref{fig2}(d) for realistic temperatures. For certain
$H_{\mathrm{ext}}$ and not too large $P_{0}$, the Kittel mode follows large
amplitude trajectories in mirror symmetric regions of the phase space (see
also SM Sec. IV \cite{SM}). With increasing $P_{0}$ the paths cross the
boundaries between attractor regions A and B. The thermal activation becomes
clear from Figure \ref{fig2}(d) that compares switching at high $T=300\,$K
(black curve), and low $T=1\,$K (purple curve), where we show single
representative trajectories in the interval $t=20-320\,\mathrm{\mu}$s. We
elaborate this stochasticity below. In SM Sec. IV \cite{SM}, we discuss the
dependence of the limit cycle trajectories on $P_{0}$ (see Fig. S4 \cite{SM}),
and analytically explain the origin of FP2 to LC transition, and its
dependence on $H_{ext}$ (see Fig. S5 \cite{SM}).

The Lindblad master equation can be solved in principle numerically exact in
number (Fock) space. With our computational facilities the Hilbert space has
to be limited to $\sim1000$, which is much too small to treat the essential
Hilbert space of a large magnet. We can reduce the Hilbert space to a
manageable size by introducing a scaling factor of the four-magnon scattering
coefficients $\mathcal{D}\rightarrow\mathcal{Q}\mathcal{D}$ with
$\mathcal{Q}\gg1$. The increased interaction preserves the topology in phase
space, but reduces the magnon amplitudes and thereby the relevant size of the
Fock space. As explained above, selecting the standing wave $\mathcal{\vec{K}%
}_{s}$, reduces the 3-mode to a 2-mode problem. We calculate up to 20 smallest
amplitude eigenvalues $\mathcal{E}\geq0$ of the r.h.s. of Eq. (\ref{eq2}), in
which the $\mathcal{E}=0$ corresponds to the ground state density matrix
$\rho_{ss}$. We visualize the steady states by the Wigner distribution
function $W(x_{0(\vec{\mathcal{K}}_{s})},p_{0(\mathcal{K}_{s})})=\int\langle
x_{0(\vec{\mathcal{K}}_{s})}-y/2|\rho_{ss,0(\vec{\mathcal{K}}_{s})}%
|x_{0(\vec{\mathcal{K}}_{s})}+y/2\rangle e^{ip_{0(\vec{\mathcal{K}}_{s})}y}dy$
of the Kittel ($\vec{\mathcal{K}}_{s}$) mode, where $|x_{0(\vec{\mathcal{K}%
}_{s})}\pm y/2\rangle$ is the position eigenstate of the Kittel ($\vec
{\mathcal{K}}_{s}$) mode, and $\rho_{ss,0(\vec{\mathcal{K}}_{s})}$ is the
density matrix after tracing out the $\vec{\mathcal{K}}_{s}$ (Kittel) mode.
The top (bottom) panels of Fig. \ref{fig3} show $W(x_{0(\vec{\mathcal{K}}%
_{s})},p_{0(\vec{\mathcal{K}}_{s})})$ for $H_{ext}=40\,$mT in each of the
`Stable', `Quantum', and `Stochastic' phases, as indicated by stars of the
same color in Fig. \ref{fig2}(a). The left and middle panels can be compared
with the classical phase space of FP1 and FP2 in Figs. \ref{fig2}(b) and (c),
respectively. The right panel of Fig. \ref{fig3} should be compared with the
limit cycle region in the classical phase space, e.g., in Fig. \ref{fig2}(c).
In the panels of Fig. \ref{fig3}, we used different scale factors
$\mathcal{Q}$ such that the distance between the extrema of $W$ is roughly the same.

The steady state classes FP1, FP2, and LC [see Fig. \ref{fig2}(a)] are
potential resources for information technologies. For a fixed input power of
$P_{0}$ the stable Ising spin can be used as a non-volatile digital memory,
while a network can operate as an Ising machine. We can assess the potential
of the device as p-bit or for quantum information by quantitative measures of
the stochasticity and entanglement derived from our semi-classical (quantum)
calculations indicated in the following by a subscript or superscript `C'
(`Q'). In `C', we solve the quantum Langevin equation of motion for Gaussian
distribution functions. In the quantum calculations, on the other hand, we
solve the master equation (\ref{eq2}) in the number (Fock) space, and the
solutions are numerically exact, but due to computational limitations we can
solve only down-scaled systems, as explained above.

\textit{Stochasticity -} The lower bound for the transition time $\tau$
between the two stable fixed points below the Suhl instability threshold
(derived in SM Sec. IIB \cite{SM}) is
\begin{align}
\mathrm{\ln}\tau &  \gtrapprox\ln\left[  \pi\left(  1+2n_{\mathrm{th}}\right)
\sqrt{(1+2\mathcal{R})/2\mathcal{R}^{2}}/2\xi_{0}\right]  +\nonumber\\
&  \left[  \left(  2\mathcal{R}+1\right)  \ln\left(  2\mathcal{R}+1\right)
-2\mathcal{R}\right]  |K_{0}|\left(  1+2n_{\mathrm{th}}\right)  /2,
\end{align}
where $\mathcal{R}=\sqrt{\mu^{2}-1}/2+1-|K_{0}|\left(  1+2n_{\mathrm{th}%
}\right)  -1$, $\mu=2P_{0}/\xi_{0}$, $K_{0}=2\mathcal{D}_{0,0,0,0}/\xi_{0}$.
Above the parametric instability threshold but below the Suhl instability
$\left(  \mu=1.1\right)  $ for typical values of $|\mathcal{D}_{0,0,0,0}%
|=1.5\times10^{-4}\,$Hz (see Fig. S5(a) \cite{SM}), $\omega_{0}/2\pi=2.5\,$GHz
[see Fig. \ref{fig1}(b)], and $T=300\,$K, this number becomes astronomically
large, $\tau\gg\exp(1.4\times10^{4})\,$s. However, by driving the system into
a limit cycle of the Kittel plus $\pm\mathcal{K}$ modes at sufficiently large
$\mu$ we find a strongly enhanced switching rate at room temperature [see Fig.
\ref{fig2}(d)]. The experimental observation of stochastic switching
\cite{Makiuchi2021} is therefore strong evidence for a parametrically driven
Suhl instability in the magnon parametron.

The telegraph noise of the Kittel mode at $T=300\,$K in Figure \ref{fig4}(a)
is caused by thermally activated random hoppings as in Fig. \ref{fig2}(d). The
calculated number of switches $N_{t}$ within $t_{e}=100\,\mathrm{\mu}$s,
averaged for several random initial conditions leads to the transition
frequencies $\mathcal{F}_{C}=N_{t}/t_{e}$ plotted in Figure \ref{fig4}(b) as a
function of $T$ (black dots). The form $l_{1}e^{-\lambda_{1}/T}+l_{2}%
e^{-\lambda_{2}/T}$, with attempt frequencies $l_{1}=32\,$Hz, $l_{2}=8.1\,$Hz,
and energy well depths $\lambda_{1}=7.98\times10^{3}\times k_{B}/2\pi\hbar
\,$GHz, $\lambda_{2}=1.5\times10^{2}\times k_{B}/2\pi\hbar\,$GHz fits the
calculations well (blue curve). We also compute the transition frequency
dependence at $T=1\,\mathrm{K}$ on $\mathcal{Q},$ a scaling number of the
four-magnon scattering coefficient $\mathcal{Q}\mathcal{D}$ that is inversely
proportional to the volume of the magnet $V_{m}.$ The red dots in Figure
\ref{fig4}(b) show that we can enhance the switching rate by either increasing
the temperature or decreasing the volume. Figure \ref{fig4}(d) shows the
dependence of $\mathcal{F}_{C}$ on $H_{\mathrm{ext}}$ and $P_{0}$ at
$T=10^{5}\,$K. Makiuchi \textit{et al}. \cite{Makiuchi2021} observed switching
frequencies $\sim0.01-0.1\,$Hz at room temperature, depending on the power
beyond a second threshold. As explained above, this is not possible without
the Suhl instability. Even though the sample in that experiment is larger than
directly accessible with our model, we can still draw conclusions from the
identical scaling for $T$ and $\mathcal{Q}$ observed in Fig. \ref{fig4}(b). By
repeating the calculations for a scaling factor $\mathcal{Q}=1/30$, we
effectively address a magnet that is $30$ times larger compared to
$\mathcal{Q}=1$. The result of $\mathcal{F}_{C}\sim0.01$ Hz at $T=300\,$K
agrees with the lower end of the experimental observations. The predicted
strong and non-monotonic dependence of $\mathcal{F}_{C}$ on $H_{\mathrm{ext}}$
in Fig. \ref{fig4}(c) also agrees with experimental findings. The substantial
enhancement of the stochasticity is due to the limit cycle dynamics with large
oscillation amplitudes which come in close vicinity of the saddle node in the
origin. Since a limit cycle broadens the distribution function when compared
to a fixed point, the thermally activated switching through the saddle node
becomes more efficient. At a fixed $H_{\mathrm{ext}}$, increasing $P_{0}$
leads to increasing LC oscillation amplitude and LC doublings (see Fig. S4
\cite{SM}), and therefore an increase in $\mathcal{F}$ is expected. Due to the
dependence of $\mathcal{D}$ coefficients on $H_{\mathrm{ext}}$ (see Fig. S5(a)
\cite{SM}), at a fixed $P_{0}$, the amplitude of the LC oscillations depends
on $H_{\mathrm{ext}}$, and has a maximum at $H_{\mathrm{ext}}\sim40\,$mT [see
SM Sec. IV and Fig. S5(d) \cite{SM}], where the maximum of both $\mathcal{F}%
_{C}$ and $\mathcal{F}_{Q}$ is observed [see Figs. \ref{fig4}(c) and (d)]. In
future work we will address a quantitative theory for large magnetic dots that
take into account perpendicular standing spin waves and three-magnon scatterings.

Next, we quantify the stochasticity from quantum calculations, i.e.
$\mathcal{F}_{Q}$. The first two eigenvalues with smallest but nonzero
$|\mathrm{\operatorname{Re}}\mathcal{E}|$ while $\mathrm{\operatorname{Im}%
}\mathcal{E}=0$, determine the tunneling frequencies (see SM Sec. IIA
\cite{SM}). One of these eigenvalues corresponds to the tunneling frequency of
the Kittel mode, $\mathcal{F}_{Q}$ \cite{Kinsler1991} [see top left panel of
Fig. \ref{fig3}(a)], as explained in SM Sec. IIA \cite{SM}. The other
corresponds to the tunneling frequency of the $\vec{\mathcal{K}}_{s}$ mode.
Below parametric instability, and below Suhl instability threshold, such
eigenvalue does not exist for either of the modes, and the $\vec{\mathcal{K}%
}_{s}$ mode, respectively.

Figure \ref{fig4}(d) shows $\mathcal{F}_{Q}$ as a function of $H_{\mathrm{ext}%
}$ for two values of $P_{0}/\xi_{0}=2.5,\,3$ that crosses both the LC and FP2
regions [see Fig. \ref{fig2}(a)]. $\mathcal{F}_{Q}$ is peaked at
$H_{\mathrm{ext}}\sim40\,$mT similar to that of $\mathcal{F}_{C}$ in Fig.
\ref{fig4}(c), and decreases sharply for $H_{\mathrm{ext}}$ in the FP2 region.
Figure \ref{fig4}(e) shows that $\mathcal{F}_{Q}$ decreases monotonically with
increasing $P_{0}$ for $H_{\mathrm{ext}}=52\,$mT where the classical steady
state does not enter the LC region. However, for $H_{\mathrm{ext}}=40\,$mT,
where the steady state changes from FP1 to FP2, and then becomes LC, by
increasing $P_{0}$, $\mathcal{F}_{Q}$ first decreases and then increases
substantially. Based on the fit in Eq. \ref{fig4}(b), for $\mathcal{Q}=10^{9}%
$, $\mathcal{F}_{C}\sim1\,$MHz, which is in the same range as expected from
$\mathcal{F}_{Q}$ in Figs. \ref{fig4}(d) and (e). The calculated
stochasticities of the Kittel mode are the same for propagating or standing
exchange waves in the large dot limit.

\begin{figure}[t]
\includegraphics[width=0.5\textwidth]{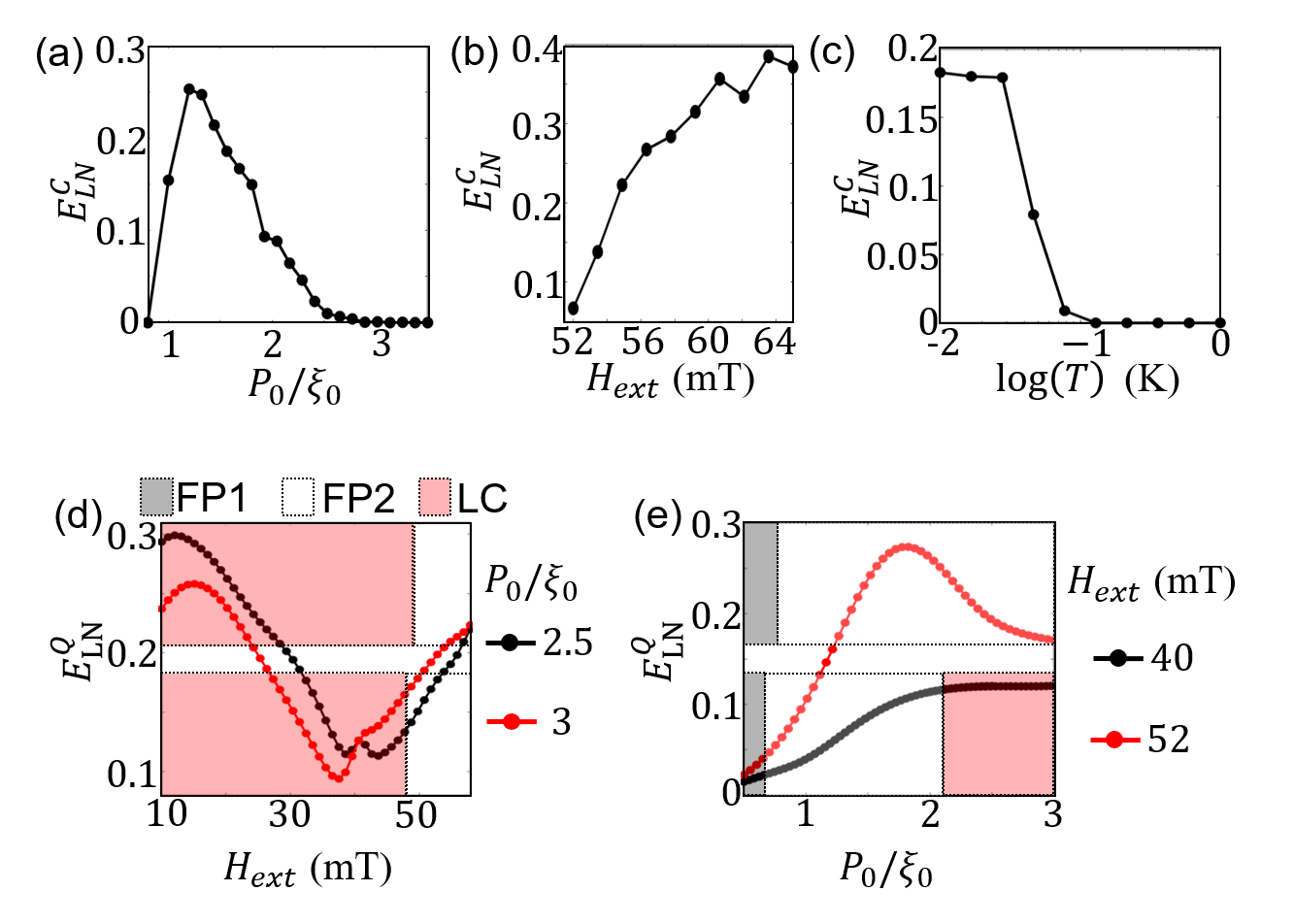}\caption{Entanglement.
(a)-(c) From semi-classical, (d)-(e) from quantum calculations. (a) Gaussian
logarithmic negativity $E_{LN}^{C}$ dependence on $P_{0}/\xi_{{0}}$ for
$H_{ext}=52\,$mT and $T=0\,$K. (b) The dependence of $E_{LN}^{C}$ on $H_{ext}$
for $P_{0}/\xi_{0}=2$ and $T=0\,$K. (c) The dependence of $E_{LN}^{C}$ on $T$
for $H_{ext}=52\,$mT and $P_{0}/\xi_{0}=2$. (d) The dependence of logarithmic
negativity $E_{LN}^{Q}$, on $H_{ext}$ for two values of $P_{0}/\xi
_{0}=2.5,\,3$. (e) The dependence of $E_{LN}^{Q}$ on $P_{0}/\xi_{0}$ for
$H_{ext}=40,\,52\,$mT. The scaling coefficient $\mathcal{Q}=5\times10^{9}$ and
$T=0\,$K in (d)-(e). The steady state class from Fig. \ref{fig2}(a)
corresponding to each point is shown by color coded rectangles, where top
(bottom) rectangles are for red (black) dots.}%
\label{fig5}%
\end{figure}\textit{Quantum entanglement -} The driven quantum state is a
many-body wave function in which the fluctuations of two types of magnons may
become quantum entangled. At the steady-state fixed points of both Kittel and
$\vec{\mathcal{K}}_{s}$ modes beyond the Suhl instability threshold, the cross
Kerr interaction $c_{0}^{\dag}c_{0}c_{\vec{\mathcal{K}}_{s}}^{\dag}%
c_{\vec{\mathcal{K}}_{s}}+\mathrm{H.c}.$ can be approximated by $\alpha
_{0}^{\ast}\alpha_{\vec{\mathcal{K}}_{s}}^{\ast}\delta c_{0}\delta
c_{\vec{\mathcal{K}}_{s}}+\mathrm{H.c}.$, where $\alpha_{0(\vec{\mathcal{K}%
}_{s})}$ is mean field value of the Kittel ($\vec{\mathcal{K}}_{s}$) mode, and
$\delta c$ indicates fluctuations. Under the conditions discussed below, this
\textquotedblleft two-mode squeezing\textquotedblright\ term leads to quantum
correlations. In the SM Sec. III \cite{SM}, we discuss the corrections by the
\textquotedblleft beam-splitter\textquotedblright\ interaction\textit{.}

The quantum correlations that characterize entanglement become apparent in the
noise statistics. The \textquotedblleft quantumness\textquotedblright\ of the
system \cite{Braunstein2005} can be measured by the mean-square fluctuations
\begin{equation}
\sigma=\left\langle \left(  \delta x_{0}-\delta x_{\vec{\mathcal{K}}_{s}%
}\right)  ^{2}+\left(  \delta p_{0}+\delta p_{\vec{\mathcal{K}}_{s}}\right)
^{2}\right\rangle =\xi_{0}\frac{n_{\mathrm{th}}+\frac{1}{2}}{2\mathfrak{g}%
+\xi_{0}}%
\end{equation}
where $\mathfrak{g}=\left\vert \mathcal{D}_{0,0,\pm\vec{\mathcal{K}},\pm
\vec{\mathcal{K}}}\alpha_{0}^{\ast}\alpha_{\vec{\mathcal{K}}_{s}}^{\ast
}\right\vert $ (see SM Sec. III \cite{SM}). In the regime $0\leq\sigma<1/2$,
the two modes are necessarily quantum-correlated or \textquotedblleft
entangled\textquotedblright. When the modes do not interact $\mathfrak{g}%
=0_{,}$ $\sigma\rightarrow n_{\mathrm{th}}+1/2.$ In general, uncorrelated and
classical states correspond to $\sigma\geq1/2.$ When $\mathfrak{g}$ becomes
large, the fluctuations and $\sigma$ vanish, which reflects commutation of the
operators for the relative positions and momenta $\left[  \delta x_{0}-\delta
x_{\vec{\mathcal{K}}_{s}},\delta p_{0}+\delta p_{\vec{\mathcal{K}}_{s}%
}\right]  =0$ in that limit. $\sigma\propto\left(  n_{\mathrm{th}}+1/2\right)
$ illustrates how increasing temperature destroys quantum correlations by
pushing the system into the classical regime $\sigma>1/2\ $irrespective of the
interactions. This allows us to estimate the experimental conditions to
observe quantum entanglement in realistic systems, see below and SM Sec. III
\cite{SM}.

For an accurate assessment of the bipartite quantum entanglement between the
Kittel and $\vec{\mathcal{K}}_{s}$ magnons fluctuations all mean-field terms
of equal order must be included, which can be done only numerically. Moreover,
$\sigma$ is not a good measure of the entanglement resource. More suitable is
the \textquotedblleft logarithmic negativity\textquotedblright\ function
$E_{LN}\ $that increases monotonically with the degree of entanglement
\cite{Vidal2002}. This parameter is measure of the negativity of the partial
transposition of the density matrix (with respect to the Kittel mode)
$\rho^{PT}=\rho_{0}^{T}\otimes\rho_{\vec{\mathcal{K}}_{s}}$ that vanishes when
the bipartite state is separable. $E_{LN}=\text{log}_{2}\left(
1+2|\mathfrak{n}|\right)  $, where $\mathfrak{n}$ is the sum of the negative
eigenvalues of $\rho^{PT}.$ $E_{LN}$ is an upper bound of the
\textquotedblleft distillable\textquotedblright\ entanglement $E_{D}$, which
is again a measure for the number of completely entangled pairs of
quasiparticles (singlets) that can be extracted from the many-body wave
function by local operations and classical communications
\cite{Braunstein2005,Bennett1996,Horodecki1996}, which are essential for e.g.
quantum teleportation and quantum key distribution
\cite{Braunstein1998,Furusawa1998,Braunstein2005,Nielsen2009}. The density
matrix of a Gaussian state, i.e. a localized state in phase space or fixed
point, is completely determined by the first and second moments of position
and momentum variables, i.e. the covariance matrix, via which $E_{LN}$ can be
readily calculated \cite{Vidal2002, Adesso2005} (see SM Sec. III \cite{SM}).
$E_{LN}=0$ for a separable bipartite state and it diverges for $\sigma=0$. In
realistic systems usually $E_{LN}<1$ \cite{Menzel2012,Palomaki2013}. Here, we
calculate the steady state covariance matrix $\mathcal{V}_{ss}$ by ensemble
averaging over 100 independent $50\,\mathrm{\mu}$s runs starting from random
initial conditions over the last $1\,\mathrm{\mu}$s. Figures \ref{fig5}(a)-(c)
summarize results of the Gaussian logarithmic negativity $E_{LN}^{C}$
calculated via $\mathcal{V}_{ss}$ for some of the FP2 cases identified
earlier. Here the superscript $C$ indicates the Gaussian assumption. In Figure
\ref{fig5}(a), we observe that $E_{LN}^{C}$ as a function of $P_{0}$, for
$H_{\mathrm{ext}}=52\,$mT and zero temperature, is zero below the Suhl
instability threshold (at $P_{0}/\xi_{0}=0.92$) and peaks at relatively small
$P_{0}$. Figure \ref{fig5}(b) shows that for a fixed $P_{0}/\xi_{0}=2$,
$E_{LN}^{C}$ increases strongly with $H_{\mathrm{ext}}$ (see Fig. S5) up to
about $\sim0.5$. According to Figure \ref{fig5}(c) $E_{LN}^{C}$ decreases with
increasing $T,$ but remains nearly constant up to $T\approx100\,$mK. In SM
Sec. III and Fig. S3 \cite{SM}, we support these observations by an analytical analysis.

The entanglement of Gaussian states may be computed by the semi-classical
approach. When nonlinearities drive the fluctuations beyond Gaussian
statistics, we have to solve the quantum master equation for the steady state
density matrix $\rho_{ss}$. The elements of $\rho_{ss}$ correspond to
$|i,j\rangle\langle i^{\prime},j^{\prime}|$, where $i$ ($i^{\prime}$) and $j$
($j^{\prime}$) refer to the $i$'th ($i^{\prime}$'th) and $j$'th ($j^{\prime}%
$'th) Fock (number) state of the Kittel ($\vec{\mathcal{K}}_{s}$) mode, and we
require the sum of the negative eigenvalues of the partially transposed
density matrix with elements $|i^{\prime},j\rangle\langle i,j^{\prime}|$.
Figure \ref{fig5}(d) shows that $E_{LN}^{Q}$ (superscript Q for quantum) is
non-monotonic in $H_{\mathrm{ext}}.$ It turns out to be minimal when the
quantum stochasticity $\mathcal{F}_{Q}$ in Figure \ref{fig4}(d) is maximal.
The phase space occupied by an LC is much larger than that of the quantum
fluctuations, hence distilling the entanglement is not feasible. Therefore,
only the entanglement in the FP2 region is useful. Figure \ref{fig5}(d) shows
that $E_{LN}^{Q}$ increases with $H_{\mathrm{ext}}$ in the FP2 region, similar
to Fig. \ref{fig5}(b). Figure \ref{fig5}(e) shows that with decreasing $P_{0}$
towards FP1 region, $E_{LN}^{Q}$ approaches zero. It can also be seen that it
has a peak in $P_{0}$ for $H_{\mathrm{ext}}=52\,$mT similar to Fig. \ref{fig5}(a).

In order to measure and distill the entanglement, both the Kittel mode and the
large wave vector magnon pair should resonantly couple to microwaves. This can
be achieved by a coplanar waveguide that is modulated with the wave length of
the wing magnons that locks the otherwise undetermined phase difference $\phi$
between the $\pm\vec{\mathcal{K}}$ pair, i.e. the nodes of the standing wave
magnon amplitude \cite{Elyasi2020}.

\textit{Conclusion -} We study the bistable nature of the Ising spin system
emulated by ferromagnetic disk parametrically excited in a microwave cavity as
a function of temperature, magnetic field, and excitation power. The Suhl
decay of the Kittel mode into a degenerate pair of magnons with large wave
vector substantially enhances the random switching between the two energy
minima of the Kittel mode parametron phase space, providing a probabilistic
bit for stochastic information processing. On the other hand, the Suhl
instability is also responsible for a finite distillable entanglement, which
is a fundamental resource for quantum information. We show that the three
regimes of operation are accessible by varying the parametric excitation power
as well as the external magnetic field. The quantum correlations in
macroscopic magnets should be observable at low but experimentally accessible
temperatures of $\sim$100 mK. We conclude that magnetic particles are
attractive building blocks for coherent Ising machines, as well as stochastic
and quantum information applications.

\textit{Acknowledgments -} We acknowledge support by JSPS KAKENHI (Nos. 19H00645 and 21K13847), and JST CREST (No. JPMJCR20C1).


\begin{thebibliography}{99}
\bibitem {SM}See supporting material [URL will be inserted by publisher] for
derivation of equations of motion from the master equation, calculation of
tunneling frequency from quantum master equation solution, analytical
discussion on tunneling frequency of a fixed point parametron, analytical
quantification of the Gaussian entanglement, demonstration of limit cycle
evolution and doublings, analytical explanation of the steady state phase
diagram including the fixed point to limit cycle bifurcation, and correlation
of limit cycle oscillation amplitude with Kittel mode parametron tunneling frequency.
\end{thebibliography}
\end{document}


\title{Theory of the Magnon Parametron: Supporting Material}
\author{Mehrdad Elyasi}
\affiliation{Institute for Materials Research, Tohoku University, Sendai 980-8577, Japan}
\author{ Eiji Saitoh}
\affiliation{Institute for Materials Research, Tohoku University, Sendai 980-8577, Japan}
\affiliation{WPI Advanced Institute for Materials Research, Tohoku University, Sendai 980-8577, Japan}
\affiliation{Center for Spintronics Research Network, Tohoku University, Sendai 980-8577, Japan}
\affiliation{Department of Applied Physics, University of Tokyo, Hongo, Tokyo 113-8656, Japan}
\affiliation{Advanced Science Research Center, Japan Atomic Energy Agency, Tokai 319-1195, Japan}
\author{Gerrit E. W. Bauer}
\affiliation{Institute for Materials Research, Tohoku University, Sendai 980-8577, Japan}
\affiliation{WPI Advanced Institute for Materials Research, Tohoku University, Sendai 980-8577, Japan}
\affiliation{Center for Spintronics Research Network, Tohoku University, Sendai 980-8577, Japan}
\affiliation{Zernike Institute for Advanced Materials, University of Groningen, 9747 AG Groningen, Netherlands}
\date{\today }

\begin{abstract}

\end{abstract}
\maketitle

The following material explains several technical details that supplement the
main text. In Sec. \ref{sec1}, we derive the Langevin equation of motion from
the Lindblad master equation. In Sec. \ref{sec2}, we address the hopping
frequency between the Ising spin states by the exact solutions of the quantum
master equation [see Sec. \ref{sec2_1}], and by estimating its value for the
Ising parametron in the classical limit [see Sec. \ref{sec2_2}]. In Sec.
\ref{sec3}, we present analytic approximations for the quantum entanglement
between the Kittel and finite-$k$ magnons that support the numerical results
in the main text. In Sec. \ref{sec4}, we discuss the limit cycles in parameter
space, explain its correlation with the hopping frequency of the Kittel mode,
and with a simple analytical approach to the steady-state phase diagram
spanned by the parametric excitation power and the applied magnetic field.

\section{\label{sec1}Equations of motion}

The starting point is the (Lindblad) equation of motion (EOM) of the density
matrix $\rho$
\begin{equation}
\dot{\rho}=-i\left[  H^{\prime},\rho\right]  +L_{d}, \label{eq_EOM_1}%
\end{equation}
where $H^{\prime}$ is the Hamiltonian for the interacting magnons in which the Kittel mode is driven by $P_{0}$
\begin{align}
H^{\prime}  &  =H_{m,L}^{\prime}+H_{m,NL}^{\prime}+\left(  P_{0}c_{0}^{\dag
}c_{0}^{\dag}+H.c.\right)  ,\nonumber\\
H_{m,L}^{\prime}  &  =\sum_{\vec{k}\in{0,\pm\mathcal{K}}}\Delta\omega_{\vec
{k}}c_{\vec{k}}^{\dag}c_{\vec{k}}\nonumber\\
H_{m,NL}^{\prime}  &  =\sum_{\vec{k}\in{0,\pm\mathcal{K}}}\left[
\mathcal{D}_{\vec{k},\vec{k},\vec{k},\vec{k}}c_{\vec{k}}^{\dag}c_{\vec{k}%
}c_{\vec{k}}^{\dag}c_{\vec{k}}+\right. \nonumber\\
&  \left.  \sum_{\vec{k}^{\prime}\in{0,\pm\mathcal{K}}}\left(  1-\delta
_{\vec{k},\vec{k}^{\prime}}\right)  \frac{1}{2}\mathcal{D}_{\vec{k},\vec
{k},\vec{k}^{\prime},\vec{k}^{\prime}}c_{\vec{k}}^{\dag}c_{\vec{k}}c_{\vec
{k}^{\prime}}^{\dag}c_{\vec{k}^{\prime}}\right]  +\nonumber\\
&  \left[  \mathcal{D}_{0,-\mathcal{K},0,\mathcal{K}}c_{0}^{\dag}c_{0}^{\dag
}c_{-\mathcal{K}}c_{\mathcal{K}}+H.c.\right]  , \label{eq_EOM_2}%
\end{align}
and%
\begin{align}
L_{d}  &  =\sum_{\vec{k}\in\{0,\pm\mathcal{K}\}}\xi_{\vec{k}}\left[
n_{\mathrm{th}}(\omega_{\vec{k}})\left(  c_{\vec{k}}\rho c_{\vec{k}}^{\dag
}+c_{\vec{k}}^{\dag}\rho c_{\vec{k}}-\rho c_{\vec{k}}c_{\vec{k}}^{\dag
}-\right. \right. \nonumber\\
& \left. \left.  c_{\vec{k}}^{\dag}c_{\vec{k}}\rho\right) +\frac{1}{2}\left(
2c_{\vec{k}}\rho c_{\vec{k}}^{\dag}-c_{\vec{k}}^{\dag}c_{\vec{k}}\rho-\rho
c_{\vec{k}}^{\dag}c_{\vec{k}}\right)  \right]  , \label{eq_EOM_3}%
\end{align}
is their dissipation into a thermal bath. Here $n_{\mathrm{th}}(\omega
_{\vec{k}})=\left(  e^{\hbar\omega_{\vec{k}}/k_{B}T}-1\right)  ^{-1}$, $k_{B}$
is the Boltzmann constant, $T$ is the bath temperature, and $\xi_{\vec{k}}$
are the dissipation rates.

We cover different regimes of the master equation (\ref{eq_EOM_1}) including
the quantum-classical crossover in the form of an equation of motion for the
Wigner quasi-probability distribution function
\begin{align}
& W(\alpha_{0},\alpha_{0}^{\ast},\alpha_{\vec{\mathcal{K}}},\alpha
_{\vec{\mathcal{K}}}^{\ast},\alpha_{-\vec{\mathcal{K}}},\alpha_{-\vec
{\mathcal{K}}}^{\ast})=\frac{1}{\pi^{2}}\int d^{2} z_{-\vec{\mathcal{K}}}\int
d^{2} z_{\vec{\mathcal{K}}}\nonumber\\
& \int d^{2} z_{0}\mathrm{tr}\left( \rho e^{iz_{0}^{\ast}c_{0}^{\dag}%
}e^{iz_{0}c_{0}}e^{iz_{\vec{\mathcal{K}}}^{\ast}c_{\vec{\mathcal{K}}}^{\dag}%
}e^{iz_{\vec{\mathcal{K}}}c_{\vec{\mathcal{K}}}}e^{iz_{-\vec{\mathcal{K}}%
}^{\ast}c_{-\vec{\mathcal{K}}}^{\dag}}e^{iz_{-\vec{\mathcal{K}}}%
c_{-\vec{\mathcal{K}}}}\right) \nonumber\\
& e^{-iz_{0}^{\ast}\alpha_{0}^{\ast}}e^{-iz_{0}\alpha_{0}}e^{-iz_{\vec
{\mathcal{K}}}^{\ast}\alpha_{\vec{\mathcal{K}}}^{\ast}}e^{-iz_{\vec
{\mathcal{K}}}\alpha_{\vec{\mathcal{K}}}}e^{-iz_{-\vec{\mathcal{K}}}^{\ast
}\alpha_{-\vec{\mathcal{K}}}^{\ast}}e^{-iz_{-\vec{\mathcal{K}}}\alpha
_{-\vec{\mathcal{K}}}},
\end{align}
where $z_{\vec{k}}$ and $\alpha_{\vec{k}}$ are complex variables, and
$|\alpha_{k}\rangle$ is the coherent state of mode $\vec{k}$. Following
textbook procedures \cite{Carmichael1999,Walls2008}
\begin{equation}
\frac{\partial W}{\partial t}=\mathcal{W}_{HO}+\mathcal{W}_{D}+\mathcal{W}%
_{PE}+\mathcal{W}_{SK}+\mathcal{W}_{CK}+\mathcal{W}_{S}, \label{eq_EOM_4}%
\end{equation}
where the contributions on the r.h.s. represent, respectively, the
non-interacting magnons
\begin{equation}
\mathcal{W}_{HO}=\sum_{k\in\{0,\pm\vec{\mathcal{K}}\}}\left[  k\Delta
\omega_{k}\frac{\partial}{\partial\alpha_{k}}\alpha_{k}+H.c.\right]  W,
\label{eq_EOM_5}%
\end{equation}
the dissipation
\begin{align}
\mathcal{W}_{D}  &  =\sum_{k\in\{0,\pm\vec{\mathcal{K}}\}}\left[  \frac
{\xi_{k}}{2}\frac{\partial}{\partial\alpha_{k}}\alpha_{k}+\frac{\xi_{k}}%
{2}\frac{\partial}{\partial\alpha_{k}^{\ast}}\alpha_{k}^{\ast}+\right.
\nonumber\\
&  \left.  \xi_{k}\left(  n_{\mathrm{th},k}+\frac{1}{2}\right)  \frac
{\partial^{2}}{\partial\alpha_{k}\partial\alpha_{k}^{\ast}}\right]  W,
\label{eq_EOM_6}%
\end{align}
the parametric excitation of the Kittel mode
\begin{align}
\mathcal{W}_{PE}=\left[  -iP_{0}\frac{\partial}{\partial\alpha_{0}}\alpha
_{0}^{\ast}+H.c.\right]  W,\label{eq_EOM_7}%
\end{align}
the self-Kerr interaction
\begin{align}
\mathcal{W}_{SK}  &  =\sum_{k\in\{0,\pm\vec{\mathcal{K}}\}}\left[
i2\mathcal{D}_{k,k,k,k}\frac{\partial}{\partial\alpha_{k}}|\alpha_{k}%
|^{2}\alpha_{k}+\right. \nonumber\\
&  \left.  \frac{1}{2}\frac{\partial^{3}}{\partial\alpha\partial\alpha
_{k}^{\ast2}}\alpha_{k}^{\ast}+H.c.\right]  W, \label{eq_EOM_8}%
\end{align}
the cross-Kerr interactions
\begin{align}
\mathcal{W}_{CK}  &  =\sum_{k\in\{0,\pm\vec{\mathcal{K}}\}}\sum_{k^{\prime}%
\in\{0,\pm\vec{\mathcal{K}}\}}k\mathcal{D}_{k,k,k^{\prime},k^{\prime}}\left(
1-\delta_{k,k^{\prime}}\right)  \left\{  \left[  {}\right.  \right.
\nonumber\\
&  \left.  \left.  \frac{\partial}{\partial\alpha_{k}}|\alpha_{k^{\prime}%
}|^{2}\alpha_{k}+\frac{1}{4}\frac{\partial^{3}}{\partial\alpha_{k}%
\partial\alpha_{k}^{\ast}\partial\alpha_{k^{\prime}}^{\ast}}\alpha_{k^{\prime
}}^{\ast}\right]  +H.c.\right\}  W, \label{eq_EOM_9}%
\end{align}
and the 4-magnon interactions that drive the Suhl instability
\begin{align}
\mathcal{W}_{S}  &  =i\mathcal{D}_{0,\mathcal{K},0,-\mathcal{K}}\left[
2\frac{\partial}{\partial\alpha_{0}}\alpha_{0}^{\ast}\alpha_{\mathcal{K}%
}\alpha_{-\mathcal{K}}-\frac{\partial}{\partial\alpha_{-\mathcal{K}}^{\ast}%
}\alpha_{0}^{\ast2}\alpha_{\mathcal{K}}+\right. \nonumber\\
&  \frac{\partial}{\partial\alpha_{\mathcal{K}}^{\ast}}\alpha_{0}^{\ast
2}\alpha_{-\mathcal{K}}-\frac{1}{4}\frac{\partial^{3}}{\partial\alpha_{0}%
^{2}\partial\alpha_{-\mathcal{K}}^{\ast}}\alpha_{\mathcal{K}}-\frac{1}{4}%
\frac{\partial^{3}}{\partial\alpha_{0}^{2}\partial\alpha_{\mathcal{K}}^{\ast}%
}\alpha_{-\mathcal{K}}+\nonumber\\
&  \left.  +\frac{1}{2}\frac{\partial^{3}}{\partial\alpha_{0}\partial
\alpha_{\mathcal{K}}^{\ast}\partial\alpha_{-\mathcal{K}}^{\ast}}\alpha
_{0}^{\ast}+H.c.\right]  W. \label{eq_EOM_10}%
\end{align}

$\partial W/\partial\alpha_{k}$ is of the order of $W$ when the latter is a peaked function such as a
Gaussian. In our system the third order derivatives in Eqs. (\ref{eq_EOM_8}%
)-(\ref{eq_EOM_10}) are neglibly small since $\left\vert \mathcal{D}\alpha
_{k}\right\vert \ll\xi_{k}\left(  n_{\mathrm{th},k}+\frac{1}{2}\right)  $.
Introducing $x_{k}=(\alpha_{k}+\alpha_{k}^{\ast})/2$ and $p_{k}=-i(\alpha
_{k}-\alpha_{k}^{\ast})/2$, the Wigner EOM of Eq. (\ref{eq_EOM_4}) reduces to
the Fokker-Planck equation (summing over repeated indices)
\begin{equation}
\dot{W}(\mathcal{X})=-\frac{\partial}{\partial\mathcal{X}_{i}}\mathcal{A}%
_{i}\left(  \mathcal{X}\right)  +\frac{1}{2}\frac{\partial^{2}}{\partial
\mathcal{X}_{i}\partial\mathcal{X}_{j}}D_{i,j},\label{eq_EOM_11}%
\end{equation}
in the variables $\mathcal{X}=\left[  x_{0},p_{0},x_{\mathcal{K}%
},p_{\mathcal{K}},x_{-\mathcal{K}},p_{-\mathcal{K}}\right]  $. $\mathcal{A}%
_{i}\left(  \mathcal{X}\right)  $ in the first term (drift) follows by
straightforward algebra from the first derivatives in Eqs. (\ref{eq_EOM_4}%
)-(\ref{eq_EOM_9}), while the second order derivatives in Eq. (\ref{eq_EOM_6})
lead to the second (diffusion) term with $D_{i,j}=\delta_{i,j}\xi_{i}\left(
n_{\mathrm{th},k}+\frac{1}{2}\right)  $. The first and second moments obtained
from the FPE can also be obtained as the solution of a set of
assoicated (Ito) stochastic differential equations \cite{Carmichael1999} in
which diffusion is obtained by Wiener increments $dW_{t}=\Xi_{i}(t)dt,$ i.e.
the differentials of Markovian Wiener processes governed by the FPE with zero
drift and unit diffusion, so
\begin{equation}
d{\mathcal{X}_{i}}=\mathcal{A}_{i}dt+\sqrt{{\xi_{i}}\left(  n_{\mathrm{th}%
,k}+\frac{1}{2}\right)  }\Xi_{i}(t)dt,\label{eq_EOM_12}%
\end{equation}
where $\langle\Xi_{i}(t)\Xi_{j}(t^{\prime})\rangle=\delta_{i,j}\delta
(t,t^{\prime})$. 

\section{\label{sec2}Hopping frequency}

As explained in the main text, the steady state of the Kittel mode above the
parametric instability threshold can be mapped on a degenerate two-level Ising
pseudo-spin that is characterized by two opposite precesssion phases. The
transition between these two states is reminiscent of the thermally activated
or quantum tunneling of the magnetization in uniformly magnetized nanoparticle
or molecular magnets \cite{Brown1963,Kovalev2011,Hayakawa2021}. In the absence
of the $\pm\vec{\mathcal{K}}$ modes, the physics of the pseudo-spin of the
Kittel mode is similar to a bistable magnetization, where the phase space of
the former is the infinite 2D plane of the Kittel mode harmonic oscillator
quasi-position and quasi-momentum and the 2D Bloch sphere in the latter. The
details of the free energy landscape on the phase spaces determine the
competition between tunneling or thermally activated hopping frequencies, but
there are some universal features as well. For example, with increasing the
parametric excitation power and in the absence of $\pm\vec{\mathcal{K}}$
modes, the pseudo spin of the Kittel mode increases while the hopping
frequency decreases [see Sec. \ref{sec2_2}], similar to the effect of
increasing the size of a nanomagnet that leads to decreasing hopping rates.
In this work, we control the Kittel mode pseudo-spin by the mixing with $\pm
\vec{\mathcal{K}}$ modes [see Sec. \ref{sec4} and Fig. 4 in the main text].

\subsection{\label{sec2_1}Number states}

We work with a finite basis set $N=N_{0}N_{\vec{\mathcal{K}}_{s}}$ , where
$N_{0(\vec{\mathcal{K}}_{s})}$ is the cutoff number of Fock states of the
$0(\vec{\mathcal{K}}_{s})$ mode. The Lindblad master Eq. (\ref{eq_EOM_1}) can
then be written as $\partial_{t}\mathcal{Z}=\mathcal{L}\mathcal{Z}$, where the
Liouvillian $\mathcal{L}$ is a $N^{2}\times N^{2}$ matrix and $\mathcal{Z}$ is
the density matrix $\rho$ rearranged into a vector with $N^{2}$ elements. The
steady state $\mathcal{Z}_{\mathrm{ss}}$ is the eigenvector of $\mathcal{L}$
with eigenvalue $\mathcal{E}_{ss}=0$. The time-dependent density matrix can be
expanded as
\begin{equation}
\mathcal{Z}_{\mathrm{phys}}(t)=\sum_{i}\mathcal{M}_{i}e^{\mathcal{E}_{i}%
t}\mathcal{Z}_{i},\label{eq_app1_1}%
\end{equation}
where the sum runs over the $N^{2}$ eigenstates, and $\mathcal{M}%
_{i}=\mathcal{Z}_{i}^{T}Z_{\mathrm{phys}}(0)$. $\forall i\neq\mathrm{ss}%
\Longrightarrow\mathrm{\operatorname{Re}}\mathcal{E}_{i}<0$ and $\lim
_{t\rightarrow\infty}\mathcal{Z}_{\mathrm{phys}}=\mathcal{Z}_{\mathrm{ss}}$.

We may model the magnon parametron by two coherent states of a harmonic
oscillator separated by a high barrier in position-momentum phase space. An
eigenvalue $\mathcal{E}_{\mathrm{tnl}}$ that satisfies
$\mathrm{\operatorname{Im}}\mathcal{E}_{\mathrm{tnl}}=0$ is then associated to
hopping. Since this rate is small compared to other fluctuations and so is
$|\mathcal{E}_{\mathrm{tnl}}|$ \cite{Kinsler1991}, we can calculate it
accurately with a small basis set. We analyze the associated tunneling
eigenvector $\mathcal{Z}_{\mathrm{tnl}}$ starting from an initial coherent state in one of the two Ising valleys $\mathcal{Z}_{\mathrm{phys}}%
(t=0)$. To leading order in the interaction $\mathcal{Z}_{\mathrm{phys}%
}(0)\approx\mathcal{Z}_{\mathrm{ss}}+\mathcal{M}_{\mathrm{tnl}}\mathcal{Z}%
_{\mathrm{tnl}}$ and $\mathcal{Z}_{\mathrm{phys}}(\infty)\approx
\mathcal{Z}_{\mathrm{ss}}$. It is convenient here to work with the
non-negative Husimi function $Q(\alpha)=\langle\alpha|\rho_{\mathrm{phys}%
}|\alpha\rangle/\pi>0$, where $|\alpha\rangle$ is a coherent state and the
$N^{2}\times N^{2}$ density matrix $\rho_{\mathrm{phys}}$ contains all
elements of the vector $\mathcal{Z}_{\mathrm{phys}}$. Figure 3 of the main
text shows that above the threshold and at long times the two valleys become
symmetrically occupied with two identical maxima of $Q(\alpha)$ representing
two mirror-symmetric coherent states. The latter corresponds to $Q(\alpha,\infty)$
for $\mathcal{Z}_{\mathrm{phys}}(\infty)=\mathcal{Z}_{ss}$, whereas the initial $Q(\alpha,0)$ corresponding
to $\mathcal{Z}_{\mathrm{phys}}(0)$ that we assumed is a coherent state at one of the valleys, has only one peak at that valley. The
decomposition $\mathcal{Z}_{\mathrm{phys}}(0)=\mathcal{Z}_{ss}+\mathcal{M}_{\mathrm{tnl}}\mathcal{Z}_{tnl}$ implies that the overlap of
$\mathcal{Z}_{\mathrm{tnl}}$ with two coherent states at the two valleys should have opposite sign in order to cancel $\mathcal{Z}_{ss}$ at one of the valleys and add to it at the other valley, for the  $Q(\alpha,0)$ to have only one peak at one of the valleys and the integration of $Q(\alpha,0)$ over the phase space to remain equivalent to that of $Q(\alpha,\infty)$.

Let us first ignore $\vec{\mathcal{K}}_{s}$ and only consider the
parametrically excited Kittel mode with self-Kerr nonlinearity. Figure
\ref{fig1}(a) shows the absolute value of the overlap $\left\vert
\mathcal{Z}_{ss}^{T}\mathcal{Z}_{\alpha}\right\vert $ between the steady state
with the elements of the coherent state density matrix $|\alpha\rangle
\langle\alpha|$ arranged into the vector $\mathcal{Z}_{\alpha}$, where
$\alpha=x_{0}+ip_{0}$. $\left\vert \mathcal{Z}_{\mathrm{tnl}}^{T}%
\mathcal{Z}_{\alpha}\right\vert =\left\vert \mathcal{Z}_{ss}^{T}%
\mathcal{Z}_{\alpha}\right\vert $, as expected. Fig. \ref{fig1}(b) shows that
$\operatorname{Re}\mathcal{Z}_{ss}^{T}\mathcal{Z}_{\alpha}$ is symmetric with
respect to $(x_{0}=0,p_{0}=0)$, while Fig. \ref{fig1}(c) indicates that
$\operatorname{Re}\mathcal{Z}_{\mathrm{tnl}}^{T}\mathcal{Z}_{\alpha}$ is
antisymmetric (and the same holds for the imaginary parts).

When the Suhl instability of the Kittel mode parametron drives the
$\vec{\mathcal{K}}_{s}$ mode, there are two tunneling eigenvalues
$\mathcal{E}_{\mathrm{tnl},1(2)}$. The Kittel mode hopping frequency is
$\mathcal{F}=-\mathcal{E}_{\mathrm{tnl},1}$ when $\left\vert \mathcal{Z}%
_{\mathrm{tnl},1}^{T}\mathcal{Z}_{\alpha}\right\vert =\left\vert
\mathcal{Z}_{ss}^{T}\mathcal{Z}_{\alpha}\right\vert $ $\forall\alpha$,
otherwise $\mathcal{F}=-\mathcal{E}_{\mathrm{tnl},2}$.

\subsection{\label{sec2_2}Kittel mode parametric oscillator}

Here, we address parametrically driven Kittel mode with finite self-Kerr
non-linearity $K_{0}$ but without interactions with other magnons, equivalent
to a Duffing oscillator. The Fokker-Planck equation for the Wigner
distribution function to leading order in the derivatives with respect to
coherent states $\alpha$ and $\alpha^{\ast}$ then reduces to
\begin{align}
\dot{W}(\alpha,\alpha^{\ast})  &  =\frac{\partial}{\partial\alpha}\left[
\alpha+i\alpha^{\ast}\left(  \mu+2K_{0}\alpha^{2}\right)  \right]
+\nonumber\\
&  \frac{1}{2}\frac{\partial^{2}}{\partial\alpha\partial\alpha^{\ast}}\left(
1+2n_{\mathrm{th}}\right)  +\mathrm{H.c.}, \label{eq_app1_2}%
\end{align}
where $K_{0}=2\mathcal{Q}\mathcal{D}_{0,0,0,0}/\xi_{0}$ and $\mu=2P_{0}%
/\xi_{0}$. With $\alpha=x_{0}+ip_{0}$,
\begin{equation}
\dot{W}(x_{0},p_{0})=\left[  -\frac{\partial}{\partial\mathcal{X}_{i}%
}\mathcal{A}_{i}+\frac{1}{2}\frac{\partial^{2}}{\partial\mathcal{X}%
_{i}\partial\mathcal{X}_{j}}D_{i,j}\right]  W(x_{0},p_{0}), \label{eq_app1_3}%
\end{equation}
$\mathcal{X}=[x_{0},p_{0}]$, $\mathcal{A}_{1}=-\left[  x_{0}-4K_{0}\left(
x_{0}^{2}+p_{0}^{2}\right)  +2\mu p_{0}\right]  $, $\mathcal{A}_{2}=-\left[
p_{0}+4K_{0}\left(  x_{0}^{2}+p_{0}^{2}\right)  +2\mu x_{0}\right]  $, and
$D_{i,j}=\delta_{i,j}\left(  1+2n_{\mathrm{th}}\right)  /2$. The maxima of $W$
are the steady-state mean-field amplitudes $\alpha_{0}=|\alpha_{0}%
|e^{i\phi_{0}}$ that follow from the first term and its conjugate (drift) of
Eq. (\ref{eq_app1_2}),
\begin{align}
i\mu\alpha_{0}^{\ast}+\alpha_{0}+2iK_{0}|\alpha_{0}|^{2}\alpha_{0}  &
=0,\nonumber\\
-i\mu\alpha_{0}+\alpha_{0}^{\ast}-2iK_{0}|\alpha_{0}|^{2}\alpha_{0}^{\ast}  &
=0. \label{eq_app1_4}%
\end{align}
Hence, $|\alpha_{0}|^{2}=\sqrt{\mu^{2}-1}|K_{0}|/2$, and $2\phi_{0}%
=\arg\left[  -K_{0}\sqrt{\mu^{2}-1}/\mu|K_{0}|+i/\mu\right]  $.

In the steady state \cite{Walls2008,Carmichael1999}
\begin{equation}
A_{i}W=\frac{1}{2}\frac{\partial}{\partial\mathcal{X}_{i}}D_{i,j}%
W.\label{eq_app1_5}%
\end{equation}
Assuming a solution of the form $W(x_{0},p_{0})=\mathcal{N}e^{-\Phi
(x_{0},p_{0})}$, where $\mathcal{N}$ is the normalizing constant
\begin{equation}
-\frac{\partial\Phi}{\partial\mathcal{X}_{i}}=2D_{i,j}^{-1}\left[  A_{j}%
-\frac{1}{2}\frac{\partial D_{j,l}}{\partial\mathcal{X}_{l}}\right]
.\label{eq_app1_6}%
\end{equation}
When $\Phi$ is a proper potential with $\partial^{2}\Phi/\partial
x_{0}\partial p_{0}=\partial^{2}\Phi/\partial p_{0}\partial x_{0}$, which
ensures that the solutions of Equations (\ref{eq_app1_6}) do not depend on the
path of integration, i.e.
\[
\Phi=-\int_{0}^{x}\frac{\partial\Phi}{\partial\mathcal{X}_{i}}d\mathcal{X}%
_{i}.
\]
Strictly speaking $\partial^{2}\Phi/\partial x_{0}\partial p_{0}=4\left[
-4K_{0}\left(  3p_{0}^{2}+x_{0}^{2}\right)  +2\mu\right]  /(1+2n_{\mathrm{th}%
})$ and $\partial^{2}\Phi/\partial p_{0}\partial x_{0}=4\left[  4K_{0}\left(
3x_{0}^{2}+p_{0}^{2}\right)  +2\mu\right]  /(1+2n_{\mathrm{th}})$ are not
equal. However, in the region between the two minima $(x_{0,m},p_{0,m})$,
$|4K_{0}|(3x_{0,m}^{2}+p_{0,m}^{2})\ll2\mu$ and $|4K_{0}|(x_{0,m}^{2}%
+3p_{0,m}^{2})\ll2\mu$. For very small $\mu>1$ above the threshold $\phi
_{0,m}\approx\pi/4$, and $(x_{0,m},p_{0,m})=(0,\pm|\alpha_{0}|)$. Therefore,
$12|K_{0}\alpha_{0}|^{2}=6\sqrt{\mu^{2}-1}\ll2\mu$, i.e., $1<\mu\ll\sqrt{9/8}%
$, approximately satisfies the potential conditions. The transition time in
this potential barrier approximation \cite{Landauer1961,Kinsler1991},
\begin{align}
\tau &  =\frac{2\pi}{\xi_{0}}\left[  \frac{\Phi_{x_{0}x_{0}}(0,0)}{\Phi
_{x_{0}x_{0}}(0,|\alpha_{0}|)\Phi_{p_{0}p_{0}}(0,|\alpha_{0}|)\Phi_{p_{0}%
p_{0}}(0,0)}\right]  ^{1/2}\times\nonumber\\
&  \exp\left[  \Phi(0,|\alpha_{0}|)-\Phi(0,0)\right]  ,\label{eq_app1_8}%
\end{align}
where $\Phi_{\mathcal{X}_{i}\mathcal{X}_{i}}=\partial^{2}\Phi/\partial
\mathcal{X}_{i}^{2}$. Hence
\begin{equation}
\tau=\frac{\pi(1+2n_{\mathrm{th}})}{2\xi_{0}}\exp\left[  \frac{\sqrt{\mu
^{2}-1}}{(1+2n_{\mathrm{th}})|K_{0}|}\right]  .\label{eq_app1_9}%
\end{equation}

\renewcommand{\thefigure}{S1} \begin{figure}[t]
\includegraphics[width=0.5\textwidth]{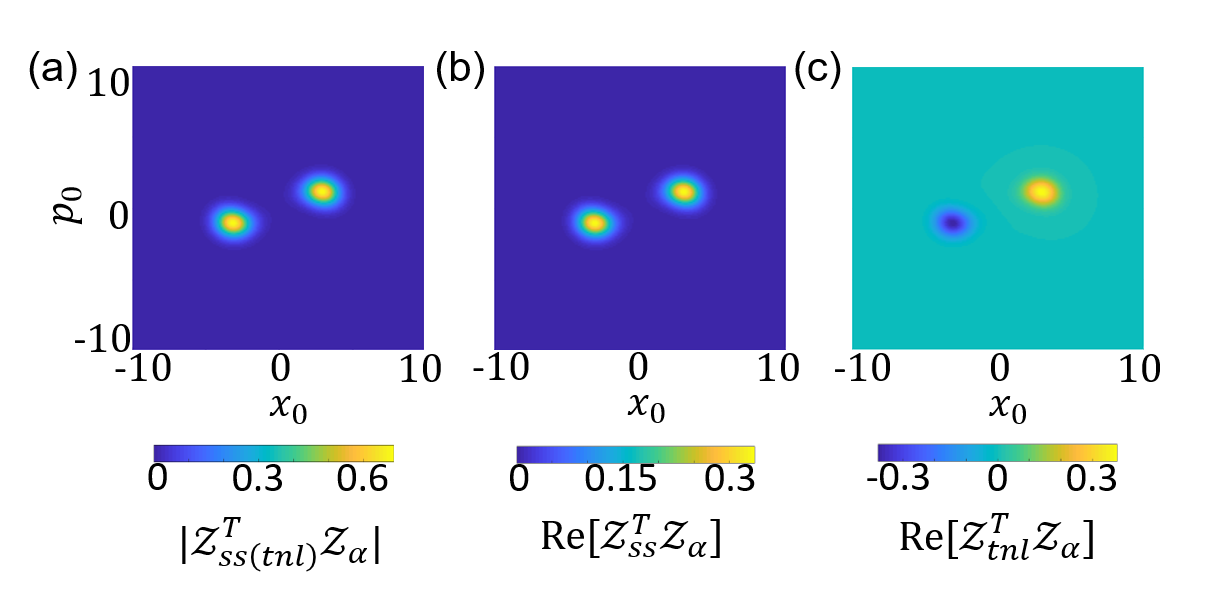}\caption{(a) The modulus
of the overlap of the Liouvillian eigenvectors $\mathcal{Z}_{ss(tnl)}$ with
the vector corresponding to the density matrix of the coherent state
$\mathcal{Z}_{\alpha}$, $\alpha=x_{0}+ip_{0}$, i.e. $\left\vert \mathcal{Z}%
_{ss\left(  tnl\right)  }^{T}\mathcal{Z}_{\alpha}\right\vert $ (b)
$\operatorname{Re}[\mathcal{Z}_{ss}^{T}\mathcal{Z}_{\alpha}]$. (c)
$\operatorname{Re}[\mathcal{Z}_{tnl}^{T}\mathcal{Z}_{\alpha}]$. }%
\label{fig1}%
\end{figure}

\renewcommand{\thefigure}{S2} \begin{figure}[t]
\includegraphics[width=0.5\textwidth]{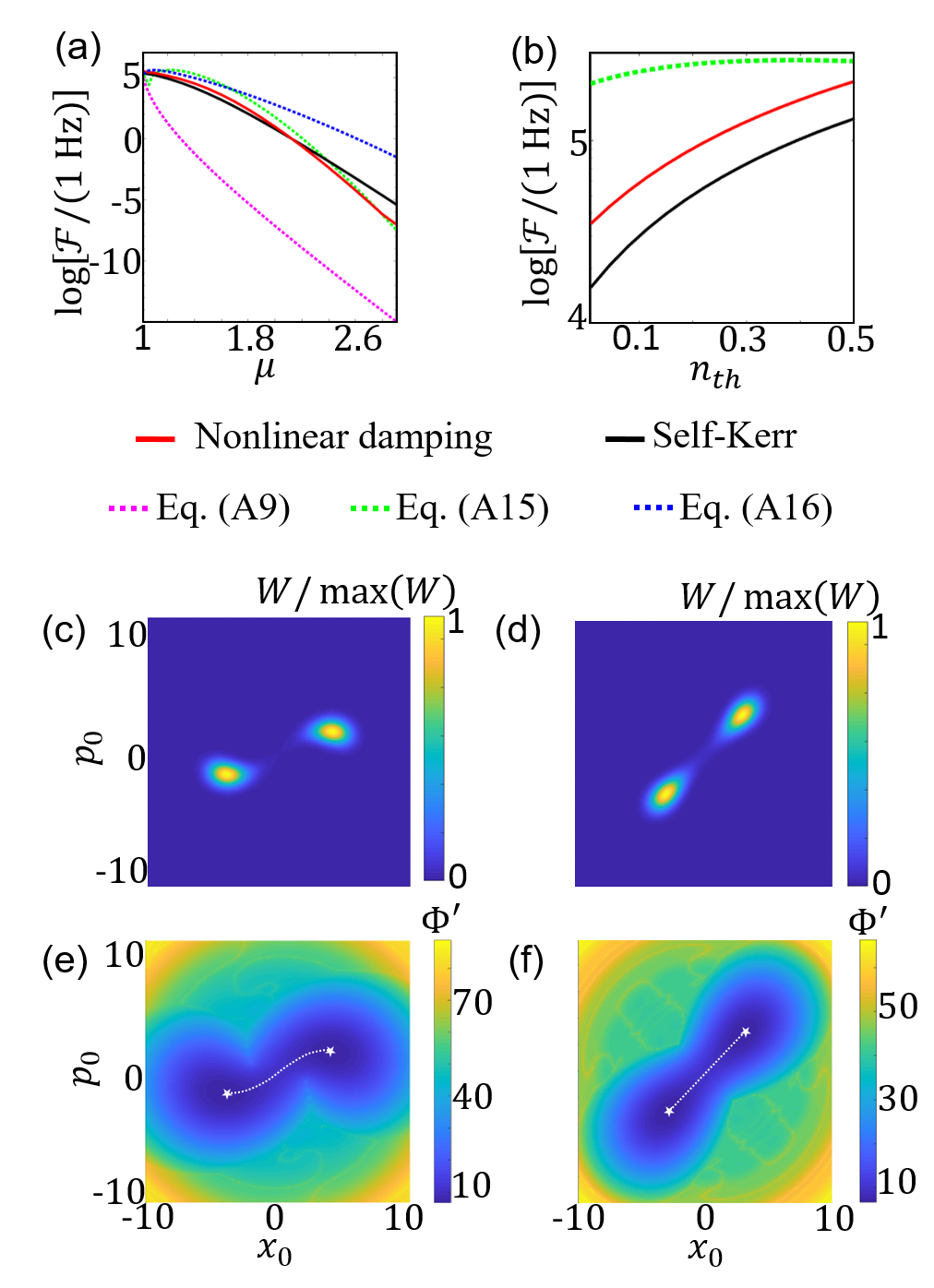}\caption{(a) The dependence of the tunneling frequency $\mathcal{F}$ on $\mu$
from quantum calculations for nonlinear damping and self-Kerr nonlinearity.
$\mathcal{F}$ from analytical equations Eq. (\ref{eq_app1_9}), Eq.
(\ref{eq_app1_15}), and Eq. (\ref{eq_app1_16}) also shown. $n_{th}=0$. (b) The
dependence of $\mathcal{F}$ on $n_{th}$ for $\mu=1.4$. In (a)-(b),
$\mathcal{Q}=5\times10^{8}$. (c) and (d) The Wigner function for self Kerr
nonlinearity and nonlinear damping, respectively, while $\mu=1.4$. (e) and (f)
The quasi-potential, $\Phi^{\prime}=-\ln W$, corresponding to (c) and (d). The
stars are the minima of $\Phi^{\prime}$ and the dashed lines are along the
minimal gradient path. In (c)-(f), $\mathcal{Q}=5\times10^{8}$. In (a)-(f),
$H_{ext}=40\,$mT.}%
\label{fig2}%
\end{figure}

Kinsler and Drummond \cite{Kinsler1991} studied the transition frequency of a
parametron in the presence of nonlinear damping rather than self-Kerr
nonlinearity, at zero temperature. Because of the similarity of the steady
states governed by either the self-Kerr interaction or the nonlinear damping
(see below), and the validity of the analytic treatment for a large $\mu$
interval in the latter case, we extend the analysis in \cite{Kinsler1991} to
nonzero temperatures through the Lindblad operator,
\begin{align}
L_{NL} &  =\xi_{NL}\left(  n_{\mathrm{th}}+1\right)  \left(  2c_{0}^{2}\rho
c_{0}^{\dag2}-c_{0}^{\dag2}c_{0}^{2}\rho-\rho c_{0}^{\dag2}c_{0}^{2}\right)
+\nonumber\\
&  \xi_{NL}n_{\mathrm{th}}\left(  2c_{0}^{\dag2}\rho c_{0}^{2}-c_{0}^{2}%
c_{0}^{\dag2}\rho-\rho c_{0}^{2}c_{0}^{\dag2}\right)  ,\label{eq_app1_10}%
\end{align}
where $\xi_{NL}$ is the nonlinear dissipation parameter. Nonlinear dissipation
should exist under parametric excitation conditions due to photon dissipation
and four-magnon scattering to thermal magnons, but in contrast to
the self-Kerr term, its value for YIG is unkown. For simplicity, we set $K_{0}=0$ and assume an imaginary parametric excitation $iP_{0}c_{0}^{\dag}c_{0}^{\dag}+H.c.$, where $P_{0}$ and thereby $\mu=2P_{0}/\xi_{0}$ are real, i.e. a phase
shift of the drive by $\pi/2$ relative to the global phase
reference. The Wigner function then solves
\begin{align}
\frac{dW}{dt} &  =\left\{  \frac{\partial}{\partial\alpha}\left[
\alpha-\alpha^{\ast}\left(  \mu-g_{NL}\alpha^{2}\right)  \right]  +\right.
\nonumber\\
&  \left.  \frac{\left(  1+2n_{\mathrm{th}}\right)  }{2}\frac{\partial^{2}%
}{\partial\alpha\partial\alpha^{\ast}}\left(  1+g_{NL}\alpha\alpha^{\ast
}\right)  +H.c.\right\}  W,\label{eq_app1_11}%
\end{align}
where $g_{NL}=4\xi_{NL}/\xi_{0}$. In terms of $(x_{0},p_{0})$, where
$\alpha=x_{0}+ip_{0}$, Eq. (\ref{eq_app1_11}) can be written in the form of
Eq. (\ref{eq_app1_3}), where $A_{1}=-x_{0}\left(  1+g_{NL}x_{0}^{2}%
-\mu\right)  $, $A_{2}=-p_{0}\left(  1+g_{NL}p_{0}^{2}+\mu\right)  $,
$D_{11}=\left(  1+2n_{\mathrm{th}}\right)  \left(  1+2g_{NL}x_{0}^{2}\right)
/4$, and $D_{22}=\left(  1+2n_{\mathrm{th}}\right)  \left(  1+2g_{NL}p_{0}%
^{2}\right)  /4$. The resulting 2D FPE does not have a potential solution,
because $\partial^{2}\Phi/\partial x_{0}\partial p_{0}\neq\partial^{2}%
\Phi/\partial p_{0}\partial x_{0}$ [see Eq. (\ref{eq_app1_6})]. However, the
states always relax towards $p_{0}=0$ because $A_{2}$ has the opposite sign to
$p_{0}$. The FPE for $p_{0}=0$ is only  a function of $x_{0}$
\cite{Kinsler1991},
\begin{align}
\frac{dW_{1\mathrm{D}}}{dt} &  =\left\{  \frac{\partial}{\partial x_{0}%
}\left[  x_{0}\left(  1+g_{NL}x_{0}^{2}-\mu\right)  \right]  +\right.
\nonumber\\
&  \left.  \frac{1}{4}\frac{\partial^{2}}{\partial x_{0}^{2}}\left[  \left(
1+2n_{\mathrm{th}}\right)  \left(  1+2g_{NL}x_{0}^{2}\right)  \right]
\right\}  .\label{eq_app1_12}%
\end{align}
All 1D FPE equations have potential solutions \cite{Walls2008}, and Eqs.
(\ref{eq_app1_6}) and (\ref{eq_app1_12}) lead to
\begin{equation}
\Phi(x_{0})=\frac{1}{\left(  1+2n_{\mathrm{th}}\right)  }\left[  x_{0}%
^{2}-\frac{2R+1}{2g_{NL}}\ln\left(  2g_{NL}x_{0}^{2}+1\right)  \right]
,\label{eq_app1_13}%
\end{equation}
where $R=\mu-g_{NL}\left(  1+2n_{\mathrm{th}}\right)  -1$. The extrema of
$\Phi_{x_{0}}$ are at $x_{0,s}=0$ (saddle point) and $x_{0,m}=\pm
\sqrt{R/g_{NL}}$ (the two minima). The transition time between the two minima
is
\begin{align}
\tau &  =\frac{2\pi}{\xi_{0}}\left[  \frac{1}{\Phi_{x_{0}x_{0}}(x_{0,m}%
)\Phi_{x_{0}x_{0}}(x_{0,s})}\right]  ^{1/2}\times\nonumber\\
&  \exp\left[  \Phi(x_{0,s})-\Phi(x_{0,m})\right]  ,\label{eq_app1_14}%
\end{align}
which with Eq. (\ref{eq_app1_13}) leads to
\begin{align}
\tau &  =\frac{\pi}{2\xi_{0}}\left[  \frac{\left(  1+2n_{\mathrm{th}}\right)
^{2}\left(  1+2R\right)  }{2R^{2}}\right]  ^{1/2}\times\nonumber\\
&  \exp\left\{  \frac{1}{2g_{NL}\left(  1+2n_{\mathrm{th}}\right)  }\left[
\left(  2R+1\right)  \ln\left(  2R+1\right)  -2R\right]  \right\}
.\label{eq_app1_15}%
\end{align}

In Fig. \ref{fig2}(a), we compare the hopping frequency $\mathcal{F}$ from a
numerical solution of the master equation corresponding to either $g_{NL}$ or
$K_{0}$ being finite, with the analytical Eqs. (\ref{eq_app1_9}) and
(\ref{eq_app1_15}), respectively, for $n_{\mathrm{th}}=0$ and a scaling
coefficient $\mathcal{Q}=5\times10^{8}$ of the four-magnon scattering
coefficients $\mathcal{D}\rightarrow\mathcal{Q}\mathcal{D}$ that reduces the
Hilbert space to a manageable size. We illustrate the results by choosing
$g_{NL}=|K_{0}|$. Figure \ref{fig2}(a) shows that Eq. (\ref{eq_app1_15})
approximately agrees with numerical calculation \cite{Kinsler1991}, whereas
Eq. (\ref{eq_app1_9}) is too small for $\mu>1$, i.e. above the parametric
pumping threshold\textit{.} because the assumption of small $\mu$ is not valid
anymore. We may improve Eq. (\ref{eq_app1_15}) by adopting a distance between
the two minima for the nonlinear damping that equals that of the self-Kerr
nonlinearity, i.e., $(\mu^{\prime}-1)/g_{NL}=\sqrt{\mu^{2}-1}/2|K_{0}|$,
\begin{align}
\tau &  =\frac{\pi}{2\xi_{0}}\left[  \frac{\left(  1+2n_{\mathrm{th}}\right)
^{2}\left(  1+2R^{\prime}\right)  }{2R^{\prime2}}\right]  ^{1/2}%
\times\nonumber\\
&  \exp\left[  \frac{1}{2g_{NL}\left(  1+2n_{\mathrm{th}}\right)  }\left\{
\left(  2R^{\prime}+1\right)  \ln\left(  2R^{\prime}+1\right)  -2R^{\prime
}\right]  \right\}  .\label{eq_app1_16}%
\end{align}
where $R^{\prime}=\sqrt{\mu^{2}-1}/2+1-g_{NL}(1+2n_{th})-1$. The blue dashed
line in Fig. \ref{fig2}(a) shows that the transition frequency from Eq.
(\ref{eq_app1_16}) now overestimates $\mathcal{F}$ for the self-Kerr
nonlinearity, and can be used as an analytical upper bound when
$n_{\mathrm{th}}=0$.

Figure \ref{fig2}(b) shows the dependence of $\mathcal{F}$ on $n_{\mathrm{th}%
}$ for $\mu=1.4$ and $\mathcal{Q}=5\times10^{8}$ for both cases of self-Kerr
and nonlinear damping. Figure \ref{fig2}(b) shows that with increasing
temperature, $\mathcal{F}$ increases as expected.
$\mathcal{F}$ according to Eq. (\ref{eq_app1_15}) monotonically increases with
temperature up to $n_{th}\approx0.5$ and is larger than the numerical
calculations [see the green dashed line in Fig. \ref{fig2}(b)]. Figure
\ref{fig2}(b) is for $\mathcal{Q}=5\times10^{8}$, which means that the
potential is $5\times10^{8}$ times shallower than the physical ($\mathcal{Q}%
=1$) case [see Eq. (\ref{eq_app1_13})]. Therefore, we can estimate that
$n_{\mathrm{th}}=0.5$ when $\mathcal{Q}=5\times10^{8}$ corresponds to
$0.5\times5\times10^{8}\sim2.5\times10^{8}$, i.e. $T\approx10^{7}\,$K when
$\mathcal{Q}=1$.
We confirm that for any $\mu$, when $\mathcal{Q}=5\times10^{8}$, and
$n_{\mathrm{th}}<0.5$, $\mathcal{F}$ from Eq. (\ref{eq_app1_16}) is larger
than the numerical calculation for the case of self-Kerr nonlinearity.
Therefore, for the physical case, Eq. (\ref{eq_app1_16}) provides a reliable
upper bound of $\mathcal{F}$ for $T<10^{7}\,$K and every value of $\mu$. 

Figures \ref{fig2}(e) and (f) show the (quasi-)potential $\Phi^{\prime}=-\ln
W$, corresponding to Figs. \ref{fig2}(c) and (d), respectively. The white
stars indicate the minima of $\Phi^{\prime}$, and the white dashed line
connects the two minima through the minimal gradient path that govern the
transition. The latter line is not straight for the self-Kerr nonlinearity in
contrast to that for nonlinear damping, supporting the validity of the 1D
assumption leading to Eq. (\ref{eq_app1_12}). Apart from the twist in the
quasi-potential of Fig. \ref{fig2}(e), it is approximately the same as that of
Fig. \ref{fig2}(f), leading to the similar scale of $\mathcal{F}$ for
self-Kerr nonlinearity and nonlinear damping. In conclusion,
while a hypothetical substantial non-linear damping should affect the numbers
but not the phenomenology and can be captured by re-scaling the Kerr constant.

\section{\label{sec3}Gaussian bipartite entanglement}

Here, we analyze the bipartite Gaussian entanglement of the Kittel mode
fluctuations with the $\vec{\mathcal{K}}_{s}$ standing wave in a magnonic dot.
\textquotedblleft Gaussian\textquotedblright\ refers to the assumption that
the fluctuations in the steady states can be enveloped by a finite ellipse and
the the second moments describe auto and cross correlations well.
\textquotedblleft Bipartite entanglement\textquotedblright\ means that the
fluctuations cannot be reproduced by a product state of Kittel mode and
$\vec{\mathcal{K}}_{s}$ fluctuations. It is caused by interactions that we
capture by the mean-field Hamiltonian with two terms called \textquotedblleft
beam splitter\textquotedblright\ and \textquotedblleft two-mode
squeezing\textquotedblright\
\begin{equation}
H_{MF}=\eta\delta c_{0}^{\dag}\delta c_{\vec{\mathcal{K}}_{s}}+\mathfrak{g}%
\delta c_{0}\delta c_{\vec{\mathcal{K}}_{s}}+\mathrm{H.c.}, \label{eq_app3_1}%
\end{equation}
where $\eta=K_{1}\alpha_{0}\alpha_{\vec{\mathcal{K}}_{s}}^{\ast}+K_{2}%
\alpha_{0}^{\ast}\alpha_{\vec{\mathcal{K}}_{s}}$ and $\mathfrak{g}=K_{1}%
\alpha_{0}^{\ast}\alpha_{\vec{\mathcal{K}}_{s}}^{\ast}$.

It is convenient to rewrite the Fokker-Planck equation of the Wigner function
Eq. (\ref{eq_app1_3}) as
\begin{equation}
\dot{W}(\vec{\mathcal{X}})=\left[  -\frac{\partial}{\partial\mathcal{X}_{i}%
}\mathcal{A}_{i,j}\mathcal{X}_{j}+\frac{1}{2}\frac{\partial^{2}}%
{\partial\mathcal{X}_{i}\partial\mathcal{X}_{j}}D_{i,j}\right]  W(\vec
{\mathcal{X}}), \label{eq_app3_2}%
\end{equation}
where $\vec{\mathcal{X}}=\left[  \delta x_{0},\delta p_{0},\delta
x_{\vec{\mathcal{K}}_{s}},\delta p_{\vec{\mathcal{K}}_{s}}\right]  $, and
$\left[  \mathcal{A}\right]  _{4\times4}$ is a mean-field drift matrix. The
diffusion matrix $D=\xi\left( 2n_{th}+1\right) /2 [I]_{4\times4}$, where $I$
is the unit matrix while
\begin{equation}
\mathcal{A}=\left[
\begin{array}
[c]{cccc}%
\xi/2 & 0 & -\mathfrak{b} & \mathfrak{a}\\
0 & \xi/2 & \mathfrak{c} & \mathfrak{d}\\
-\mathfrak{d} & \mathfrak{a} & \xi/2 & 0\\
\mathfrak{c} & \mathfrak{b} & 0 & \xi/2
\end{array}
\right]  , \label{eq_app3_3}%
\end{equation}
where $\mathfrak{a}=\mathrm{\operatorname{Re}}\left(  \mathfrak{g}%
\ -\eta\right)  $, $\mathfrak{b}=\operatorname{Im}\left(  \mathfrak{g}%
+\eta\right)  $, $\mathfrak{c}=\mathrm{\operatorname{Re}}\left(
\mathfrak{g}\ +\eta\right)  $, and $\mathfrak{d}=\operatorname{Im}\left(
\mathfrak{g}-\eta\right)  $. The steady state covariance matrix with elements
$\langle\mathcal{X}_{i}\mathcal{X}_{j}\rangle$,
\begin{equation}
\mathcal{V}_{ss}=\left[
\begin{array}
[c]{cc}%
\Upsilon & \Xi\\
\Xi^{T} & \Upsilon^{\prime}%
\end{array}
\right]  \label{eq_app3_5}%
\end{equation}
solves
\begin{equation}
\mathcal{A}\mathcal{V}_{ss}+\mathcal{V}_{ss}\mathcal{A}=-D. \label{eq_app3_4}%
\end{equation}

The distillable bipartite entanglement is bounded from above by the
logarithmic negativity \cite{Vidal2002,Braunstein2005,Adesso2005},
\begin{align}
E_{LN}  &  =\max\left[  0,-\ln\left(  \zeta\right)  \right]  \label{eq_app3_6}%
\\
\zeta^{2}  &  =\left[  f(\mathcal{V}_{ss})-\sqrt{f^{2}(\mathcal{V}%
_{ss})-4|\mathcal{V}_{ss}|}\right]  , \label{eq_app3_7}%
\end{align}
where $f(\mathcal{V}_{ss})=\det \Upsilon+\det \Upsilon^{\prime}-2\det\Xi\ $and
$\det$ indicates the determinant of a matrix.

\renewcommand{\thefigure}{S3} \begin{figure}[t]
\includegraphics[width=0.5\textwidth]{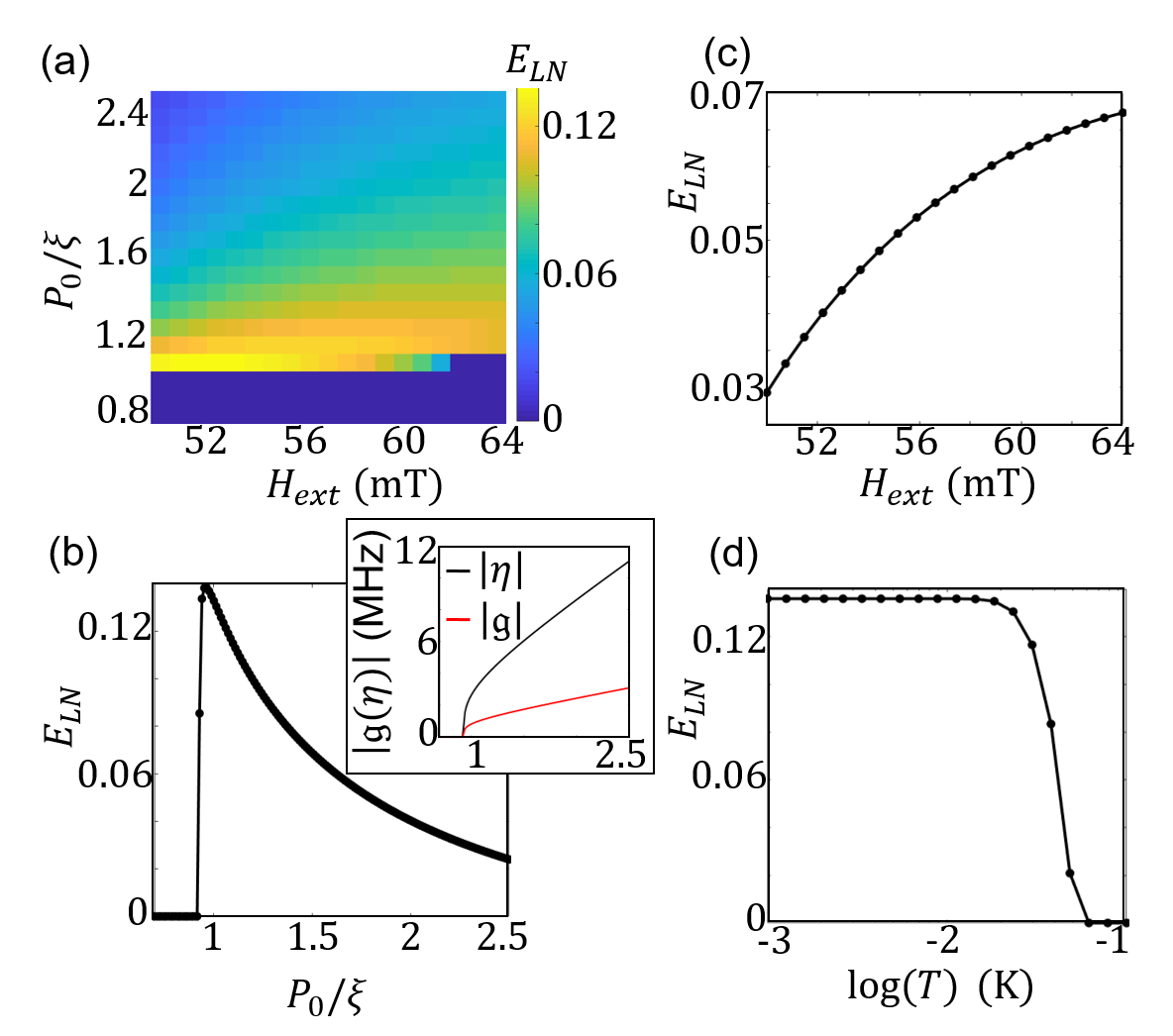}\caption{Results from an
analytical model for Gaussian magnon entanglement. (a) The logarithmic
negativity $E_{LN}$ as a function of $H_{\mathrm{ext}}$ and $P_{0}/\xi$. Panels (b)-(d) are line cuts out of
(a). (b) Dependence of $E_{LN}$ and $\left\vert \mathfrak{g}(\eta)\right\vert
$ (inset) on $P_{0}/\xi$ at $H_{ext}=52\,$mT. (c) $E_{LN}$ dependence on
$H_{ext}$ at $P_{0}/\xi=2$. (d) $E_{LN}$ dependence on $T$ at $H_{ext}=50\,$mT
and $P_{0}/\xi=1$.}%
\label{fig3}%
\end{figure}

Assuming without loss of generality $\operatorname{Im}\mathfrak{g}%
=\operatorname{Im}\eta=0$, the solution of Eq. (\ref{eq_app3_4}) are
\begin{align}
\mathcal{V}_{ss,1,1}  &  =\frac{2n\left(  2\mathfrak{a}\mathfrak{c}-\xi
^{2}/4-2\mathfrak{a}^{2}\right)  }{\xi\left(  4\mathfrak{a}\mathfrak{c}%
-\xi^{2}/4\right)  },\nonumber\\
\mathcal{V}_{ss,1(4),4(1)}  &  =\frac{n\left(  \mathfrak{a}+\mathfrak{c}%
\right)  }{\left(  4\mathfrak{a}\mathfrak{c}-\xi^{2}/4\right)  },\nonumber\\
\mathcal{V}_{ss,2,2}  &  =\frac{2n\left(  2\mathfrak{a}\mathfrak{c}-\xi
^{2}/4-2\mathfrak{c}^{2}\right)  }{\xi\left(  4\mathfrak{a}\mathfrak{c}%
-\xi^{2}/4\right)  },\nonumber\\
\mathcal{V}_{ss,2(3),3(2)}  &  =\mathcal{V}_{ss,1(4),4(1)}, \label{eq_app3_8}%
\end{align}
and lead to
\begin{equation}
\zeta^{2}=\frac{n^{2}\xi^{2}}{4\Omega}\left[  \mathcal{J}+\xi^{2}\left(
\mathfrak{a}+\mathfrak{c}\right)  /4-\xi\left(  \mathfrak{a}+\mathfrak{c}%
\right)  \sqrt{\mathcal{J}}\right]  , \label{eq_app3_9}%
\end{equation}
where $n=\xi(2n_{th}+1)/2$, $\mathcal{J}=\left(  2\mathfrak{a}\mathfrak{c}%
-\xi^{2}/4-2\mathfrak{a}^{2}\right)  \left(  2\mathfrak{a}\mathfrak{c}-\xi
^{2}/4-2\mathfrak{c}^{2}\right)  $, and $\Omega=\xi^{4}\left(  4\mathfrak{a}%
\mathfrak{c}-\xi^{2}/4\right)  ^{2}/16$.

We proceed by the rather rough approximation that beyond the Suhl instability
the Kittel mode mean-field $\alpha_{0}=|\alpha_{0}|e^{i\phi_{0}}$ is constant
with
\begin{align}
|\alpha_{0}|^{2} &  =\frac{1}{2|\mathcal{D}_{0,0,0,0}|}\sqrt{P_{0}^{2}-\left(
\frac{\xi}{2}\right)  ^{2}},\nonumber\\
\phi_{0} &  =-\arg\left[  \frac{-i}{P_{0}}\left(  \frac{\xi}{2}+2i\mathcal{D}%
_{0,0,0,0}|\alpha_{0}|^{2}\right)  \right]  /2,\label{eq_app3_10}%
\end{align}
which is the steady state of the self Hamiltonian $(P_{0}c_{0}^{\dag}%
c_{0}^{\dag}+\mathrm{H.c.})+\mathcal{D}_{0,0,0,0}c_{0}^{\dag}c_{0}c_{0}^{\dag
}c_{0}$. The mean field $\alpha_{\vec{\mathcal{K}}_{s}}$ is then the steady
state of the effective Hamiltonian $\Delta^{\prime}\omega_{\vec{\mathcal{K}%
}_{s}}c_{\vec{\mathcal{K}}_{s}}^{\dag}c_{\vec{\mathcal{K}}_{s}}+(K_{2}%
\alpha_{0}^{2}c_{\vec{\mathcal{K}}_{s}}^{\dag}c_{\vec{\mathcal{K}}_{s}}^{\dag
}+H.c.)+K_{5}c_{\vec{\mathcal{K}}_{s}}^{\dag}c_{\vec{\mathcal{K}}_{s}}%
c_{\vec{\mathcal{K}}_{s}}^{\dag}c_{\vec{\mathcal{K}}_{s}}$, where
$\Delta^{\prime}\omega_{\vec{\mathcal{K}}_{s}}=\Delta\omega_{\vec{\mathcal{K}%
}_{s}}+K_{1}|\alpha_{0}|^{2}$. $K_{1}=\mathcal{D}_{0,0,\pm\vec{\mathcal{K}%
},\pm\vec{\mathcal{K}}}$, $K_{2}=\mathcal{D}_{0,\vec{\mathcal{K}}%
,0,-\vec{\mathcal{K}}}$, $K_{5}=2K_{3}+K_{4}$, $K_{3}=\mathcal{D}_{\pm
\vec{\mathcal{K}},\pm\vec{\mathcal{K}},\pm\vec{\mathcal{K}},\pm\vec
{\mathcal{K}}}$, and $K_{4}=\mathcal{D}_{\pm\vec{\mathcal{K}},\pm
\vec{\mathcal{K}},\mp\vec{\mathcal{K}},\mp\vec{\mathcal{K}}}$ [cf. Eq. (4) in
the main text to determine],
\begin{align}
\alpha_{\vec{\mathcal{K}}_{s}} &  =|\alpha_{\vec{\mathcal{K}}_{s}}%
|e^{i\phi_{\vec{\mathcal{K}}_{s}}},\nonumber\\
|\alpha_{\vec{\mathcal{K}}_{s}}|^{2} &  =\frac{1}{2|K_{5}|}\sqrt{K_{2}%
^{2}|\alpha_{0}|^{4}-\left(  \frac{\xi}{2}\right)  ^{2}}-\Delta^{\prime}%
\omega_{\vec{\mathcal{K}}_{s}},\nonumber\\
\phi_{\vec{\mathcal{K}}_{s}} &  =\arg\left[  \frac{-i}{2\alpha_{0}^{2}}\left(
-\frac{\xi}{2}-2iK_{5}|\alpha_{\vec{\mathcal{K}}_{s}}|^{2}-i\Delta^{\prime
}\omega_{\vec{\mathcal{K}}_{s}}\right)  \right]  /2.\label{eq_app3_11}%
\end{align}

Figure \ref{fig3}(a) shows the resulting $E_{LN}$ as a function of $P_{0}/\xi$
and $H_{\mathrm{ext}}$. At a fixed $H_{\mathrm{ext}}$, $E_{LN}$ jumps to a
nonzero value just above the Suhl instability threshold to subsequently
decrease with increasing $P_{0}/\xi$ [see also Fig. \ref{fig3}(b)]. For
$P_{0}/\xi>1.4$, $E_{LN}$ increases with $H_{\mathrm{ext}}$ [see also Fig.
\ref{fig3}(c)]. Figure \ref{fig3}(d) shows that $E_{LN}$ decreases with $T$
and drops to $\sim0.01$ at around $T\sim100\,$mK. The good agreement with the
numerical results in Fig. 5 of the main text confirm the validity of the model assumptions.

The self-consistent mean-fields $\left\vert \mathfrak{g}\right\vert $ and
$\left\vert \eta\right\vert $ monotonically increase with $P_{0}$, and
eventually $\left\vert \eta\right\vert \gg\left\vert \mathfrak{g}\right\vert $
[see the inset of Fig. \ref{fig3}(b)]. For large $P_{0}$, $\mathcal{J}%
\approx\Omega\approx\left(  4\eta^{2}+\xi^{2}/4\right)  ^{2}$ [defined after
Eq. (\ref{eq_app3_9})] and
\begin{equation}
\zeta^{2}=\left(  2n_{\mathrm{th}}+1\right)  \left[  1+\frac{\mathfrak{g}%
^{2}\xi^{2}/4-\xi\mathfrak{g}\left(  4\eta^{2}+\xi^{2}/4\right)  }{\left(
4\eta^{2}+\xi^{2}/4\right)  ^{2}}\right]  ,\label{eq_app3_12}%
\end{equation}
leading to $\lim_{\eta\rightarrow\infty}\zeta^{2}=1$ and $\lim_{P_{0}%
\rightarrow\infty}E_{LN}=0$. The entanglement vanishes for large $P_{0}$ [see
\ref{fig3}(b)], as it does in the numerical calculations [see Fig. 5(a) of the
main text].

Just above the Suhl instability threshold, $\xi/2\gg\mathfrak{g},\eta$, i.e.,
$\mathcal{J}\approx\Omega\approx\xi^{4}/16$, and
\begin{equation}
\zeta^{2}=\left(  2n_{\mathrm{th}}+1\right)  \frac{\xi^{4}+4\mathfrak{g}%
^{2}\xi^{2}-4\mathfrak{g}\xi^{3}}{\xi^{4}}.\label{eq_app3_13}%
\end{equation}
while below it, $\mathfrak{g}=0$, and Eq. (\ref{eq_app3_9}) leads to
$\zeta^{2}=1$, i.e., $E_{LN}=0$. Above the Suhl instability, $\mathfrak{g}$
increases with $P_{0}$, and according to Eq. (\ref{eq_app3_13}), just above
thee threshold, $\zeta^{2}$ drops below 1 and decreases, i.e., according to
Eq. (\ref{eq_app3_6}) $E_{LN}$ becomes nonzero and increases [see Figs.
\ref{fig3}(b) and 5(a) of the main text].

According to Eq. (\ref{eq_app3_12}), $\zeta^{2}$ increases with
$n_{\mathrm{th}}$, i.e., $E_{LN}$ decreases as $-\ln[(2n_{\mathrm{th}%
}+1)]+E_{LN}(T=0)$ [see Fig. \ref{fig3}(d)]. A magnetic field freezes
$n_{\mathrm{th}}$ out and $E_{LN}(T)\rightarrow E_{LN}(T=0)$.

In spite of the approximations, the analytical model appears to explain the
qualitative physics well, but we emphasize that they cannot replace the
accurate numerical calculations presented in Figs. 5(a)-(c) of the main text.

\section{\label{sec4}Suhl instability and bifurcation of the limit cycles}

\renewcommand{\thefigure}{S4} \begin{figure}[t]
\includegraphics[width=0.5\textwidth]{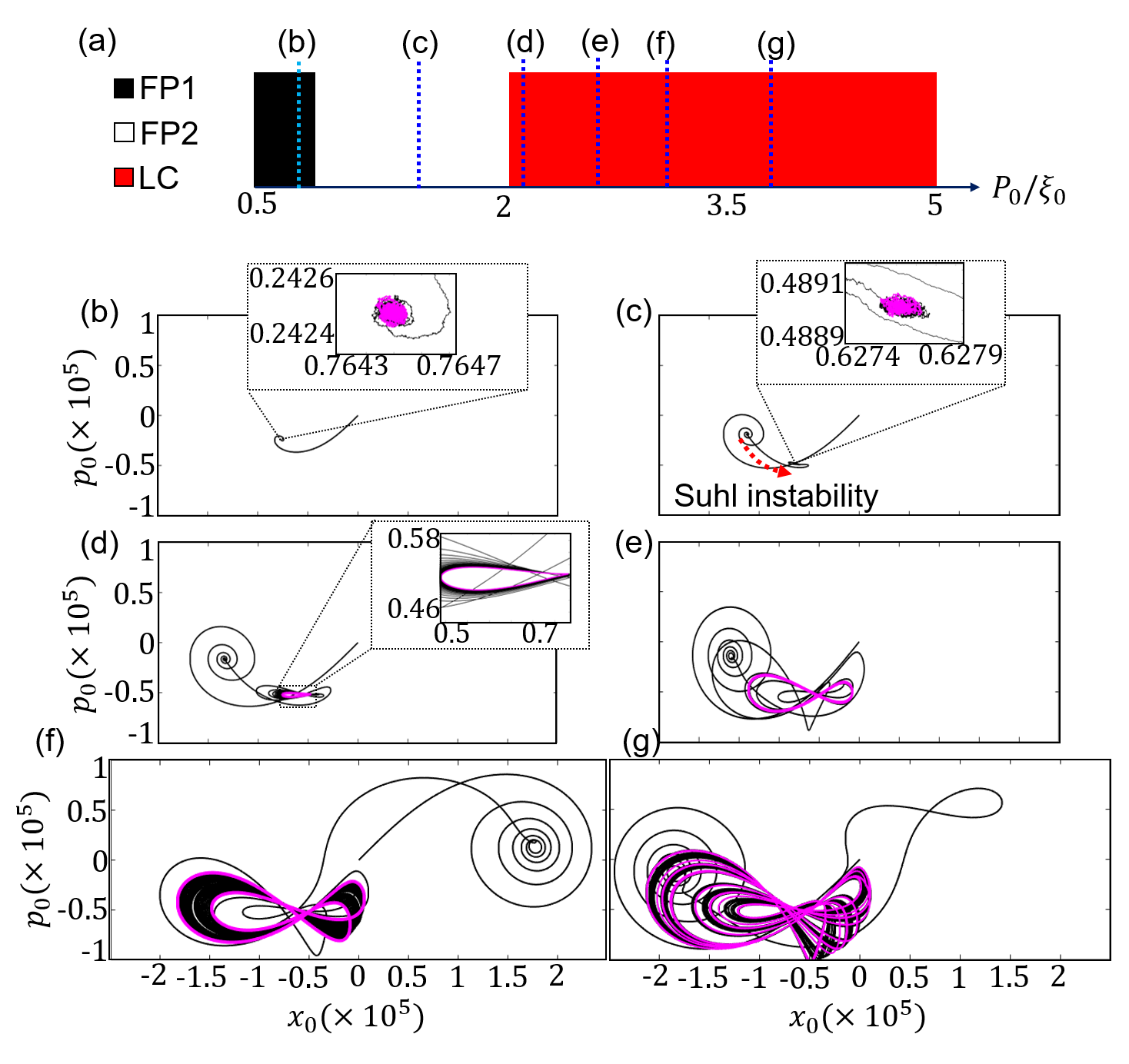}\caption{(a) The
dynamical steady states of the magnetic dot calculated as a function of
$P_{0}/\xi_{0}$, for $H_{ext}=40\,$mT. The blue dashed lines indicate
$P_{0}/\xi_{0}$ that corresponding to the time traces plotted in panels
(b)-(g). (b) $P_{0}/\xi_{0}=0.86$, FP1. (c) $P_{0}/\xi_{0}=1.51$, FP2. The red
dashed arrow indicates the Suhl instability due to parametric pumping. (d)
$P_{0}/\xi_{0}=2.06$, LC. (e) $P_{0}/\xi_{0}=2.34$, LC. (f) $P_{0}/\xi
_{0}=2.89$, LC. (g) $P_{0}/\xi_{0}=3.81$, LC. In (b)-(g), the black
trajectories are from $t=0$ to $t=40\,\mu$s, and the purple trajectories are
from $t=30\,\mu$s to $t=40\,\mu$s. The insets in (b)-(d) are zooms of the main
panels.}%
\label{fig4}%
\end{figure}

\renewcommand{\thefigure}{S5} \begin{figure}[t]
\includegraphics[width=0.5\textwidth]{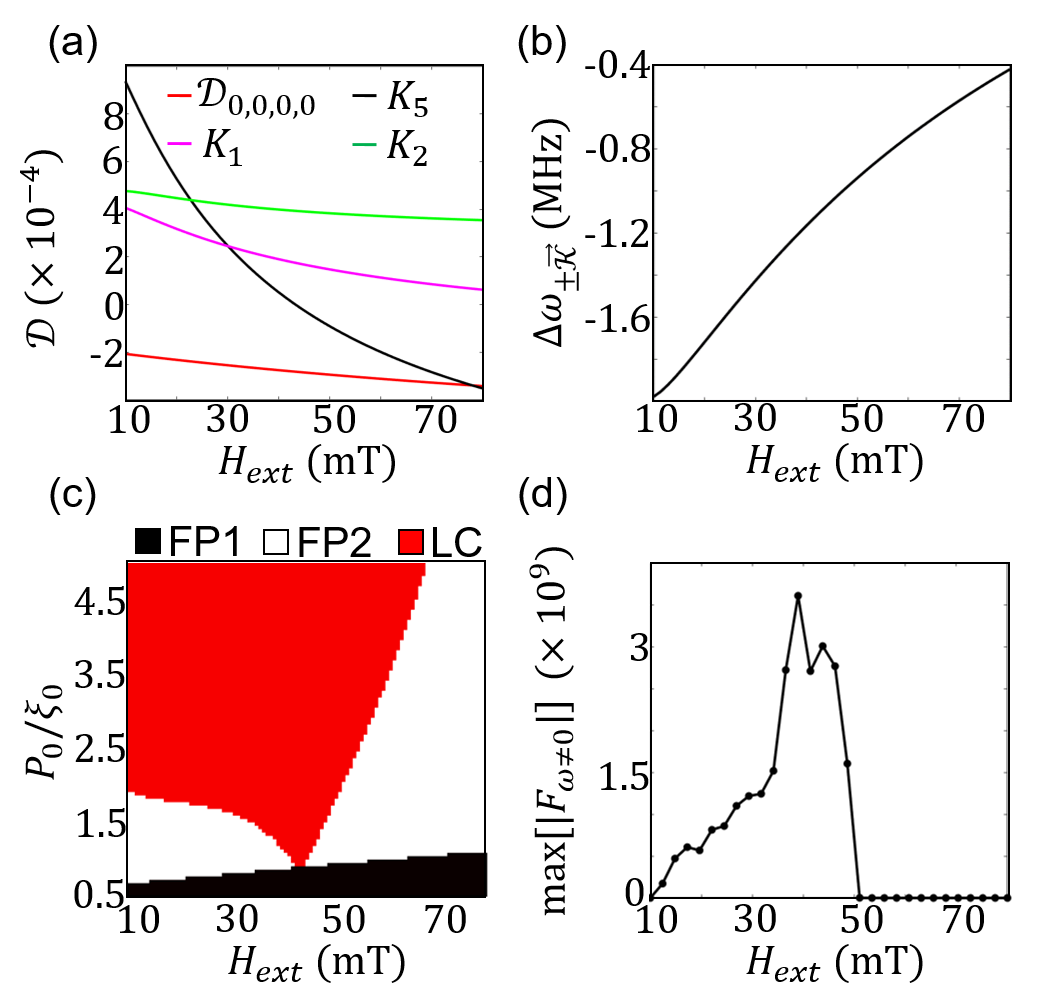}\caption{(a) The
dependence of four-magnon scattering coefficients relevant to the dynamics on
$H_{ext}$. (b) The dependence of $\Delta\omega_{\pm\vec{\mathcal{K}}}$ on
$H_{ext}$. (c) An analytical phase diagram [see Fig. 2(a) of the main text].
(d) The $H_{ext}$ dependence of the peak of Fourier transform of $x_{0}(t)$,
$F_{\omega}$, at any $\omega\neq0$, for $P_{0}/\xi_{0}=3.5$ [see Figs. 2(a)
and 4(c) of the main text].}%
\label{fig5}%
\end{figure}

In Sec. \ref{sec2_2}, we discussed the hopping frequency $\mathcal{F}$ of the
Kittel mode parametron in the absence of interactions with $\pm\vec
{\mathcal{K}}$ pair. We showed in the main text that for $P_{0}$ above the
Suhl instability that leads to the decay into finite momentum magnon pairs the
transition rate increases substantially when the steady state is a limit cycle
(LC). Here, we discuss the LC trajectories as a function of $P_{0}$ up to just
below the jump in $\mathcal{F}$. Figure \ref{fig4}(a) presents our results for
the steady states as a function of $P_{0}/\xi_{0}$ (see also Fig. 2 of the
main text), for $H_{\mathrm{ext}}=40\,$mT, and indicates the respective value
of $P_{0}/\xi_{0}$ corresponding to Figs. \ref{fig4}(b)-(g) in increasing
order. In Figs. \ref{fig4}(b)-(g), we assumed $T=0$ and an initial random
$(x_{0},p_{0})$ near the origin. The black trajectories are from $t=0$ to
$t=40\,\mu$s, whereas the purple ones are the steady state trajectories from
$t=30\,\mathrm{\mu}$s to $t=40\,\mathrm{\mu}$s. Figure \ref{fig4}(b) is the
fixed point FP1 steady state below the Suhl instability. Figure \ref{fig4}(c)
is the FP2 fixed point steady state generated by the Suhl instability. The
Ising spin states of FP1 and FP2 are extremely stable, with astronomically
small hopping frequencies.

Figure \ref{fig4}(d) correspond to a microwave power just above the
bifurcation to a LC steady state. Initially, the LC has a small amplitude that
increases with $P_{0}/\xi_{0}$ to develop a pronounced butterfly shape [see
Fig. \ref{fig4}(e)]. For even larger $P_{0}/\xi_{0}$, the limit cycle
bifurcates. In the steady state in Fig. \ref{fig4}(f) we observe 2 LC
butterflies and 4 in Fig. \ref{fig4}(g). The increase in the LC radius and its
multiplying indicates a shallower potential trench that confines the dynamics
and a much broader probability distribution that is definitely not Gaussian. A shallower potential well
or trench implies an increase in the hopping over the saddle point between the
two Ising spin states. 

Next, we discuss field dependence of the steady state phase diagram of Fig.
2(a) of the main text and the associate Ising spin flip rates. At a fixed
$P_{0}$, the steady states are governed by the four magnon scattering
coefficients $\mathcal{D}$, and $\Delta\omega_{\pm\vec{\mathcal{K}}}$, that in
turn depend on $H_{\mathrm{ext}}$. Figure \ref{fig5}(a) shows the dependence
of $\mathcal{D}_{0,0,0,0}$, $K_{1}=\mathcal{D}_{0,0,\pm\vec{\mathcal{K}}%
,\pm\vec{\mathcal{K}}}$, $K_{2}=\mathcal{D}_{0,\vec{\mathcal{K}}%
,0,-\vec{\mathcal{K}}}$, and $K_{5}=2K_{3}+K_{4}$ on $H_{\mathrm{ext}}$, where
$K_{3}=\mathcal{D}_{\pm\vec{\mathcal{K}},\pm\vec{\mathcal{K}},\pm
\vec{\mathcal{K}},\pm\vec{\mathcal{K}}}$ and $K_{4}=\mathcal{D}_{\pm
\vec{\mathcal{K}},\pm\vec{\mathcal{K}},\mp\vec{\mathcal{K}},\mp\vec
{\mathcal{K}}}$. Figure \ref{fig5}(b) shows the dependence of $\Delta
\omega_{\pm\vec{\mathcal{K}}}$ on $H_{ext}$. As we discussed in the main text,
the boundary of FP1 and FP2 [the green line in Fig. 2(a) of the main text] is
the Suhl instability threshold. According to Eq. (4) of the main text
\begin{align}
|\alpha_{0,\mathrm{Suhl}}|^{2} &  =\frac{1}{K_{2}^{2}-K_{1}^{2}}\times\left(
\Delta\omega_{\pm\vec{\mathcal{K}}}K_{1}+\right.  \nonumber\\
&  \left.  \sqrt{{\Delta\omega_{\pm\vec{\mathcal{K}}}}^{2}K_{1}^{2}+(K_{2}%
^{2}-K_{1}^{2})\times(\xi_{0}^{2}/4+{\Delta\omega_{\pm\vec{\mathcal{K}}}}%
^{2})}\right)  .\label{eq_app2_1}%
\end{align}
For $\xi_{0}=5\,$MHz, $|\alpha_{0,\mathrm{Suhl}}|^{2}\approx\xi_{0}%
/2\sqrt{K_{2}^{2}-K_{1}^{2}}$ decreases by a factor of $\sim1/1.3$ when
increasing $H_{\mathrm{ext}}$ from $10\,$mT to $80\,$mT while $|\alpha
_{0}|^{2}\propto1/|\mathcal{D}_{0,0,0,0}|$ decreases by a factor of
$\sim1/1.7$. The $P_{0}$ needed to drive the Suhl instability therefore
increases slightly by a factor of $\sim1.7/1.3$ in the field interval, as does
the green line in Fig. 2(a) of the main text.

We can estimate the boundary between FP2 and LC phases by the same
approximation that led to $\alpha_{0}$ and $\alpha_{\vec{\mathcal{K}}_{s}}$ in
Eqs. (\ref{eq_app3_11}) and (\ref{eq_app3_12}) above the threshold. The
transition for a limit cycle by the mean field term $K_{1}\alpha_{0}^{\ast
}\alpha_{\vec{\mathcal{K}_{s}}}^{\ast}\delta c_{0}\delta c_{\vec
{\mathcal{K}_{s}}}+\mathrm{H.c.}$ requires that
\begin{equation}
\left\vert K_{1}\alpha_{0,LC}\alpha_{\vec{\mathcal{K}_{s}},LC}\right\vert
=\xi_{0}/2, \label{eq_app2_2}%
\end{equation}
where $\alpha_{0,LC}$ and $\alpha_{\vec{\mathcal{K}_{s}},LC}$ are the mean
fields at the FP2 to LC transition. For the same magnetic field interval and
fixed $P_{0}$, $|\alpha_{0}|^{2}\propto1/|\mathcal{D}_{0,0,0,0}|$ decreases by
a factor of $\sim1/1.7$. In $|\alpha_{{\vec{\mathcal{K}}_{s}}}|^{2}%
\propto(K_{2}+K_{1})|\alpha_{0}|^{2}/\left\vert K_{5}\right\vert \propto
(K_{2}+K_{1})/|K_{5}||\mathcal{D}_{0,0,0,0}|,$ $(K_{2}+K_{1})/\mathcal{D}%
_{0,0,0,0}|$ decreases by a factor of $\sim0.3$ while $|K_{5}|$ decreases to
zero at around $40\,$mT, and then increases to $\sim0.5$ times of the initial
value, at $80\,$mT. $K_{1}$ monotonically decreases by a factor of $\sim1/8$
with increasing $H_{ext}$ in the same range. Therefore, $P_{0}$ corresponding
to $|K_{1}\alpha_{0,LC}\alpha_{\vec{\mathcal{K}_{s}},LC}|$ has a minimum at
$H_{\mathrm{ext}}\sim40\,$mT, and is larger by a factor of $\sim8$ for
$H_{\mathrm{ext}}=80\,$mT than $H_{\mathrm{ext}}=10\,$mT. Fig. \ref{fig5}(c)
shows the boundaries between FP2, FP1 and LC as calculated from Eqs.
(\ref{eq_app2_2}) and (\ref{eq_app3_11})-(\ref{eq_app3_12}) and captures the
main features of the numerically exact phase diagram of Fig. 2(a) of the main
text. The differences such as the strong drop of the FP2%
$\vert$%
LC boundary at intermediate fields are a consequence of the rough mean-field approximation.

We expect that for fixed $P_{0}$ beyond the LC transition the limit cycle
oscillation amplitudes would be enhanced at $H_{\mathrm{ext}}\sim40\,$mT. This
is indeed consistent with the Ising spin hopping rate $\mathcal{F}$ in Figs.
4(c) and (d) of the main text. $F_{\omega},$ the Fourier transform of
$x_{0}(t)$ may help to elucidate the correlation between the LC oscillation
amplitude and $\mathcal{F}$ since a peaked $F_{\omega}$ indicates that the
steady state is a limit cycle. Figure \ref{fig5}(d) shows the dependence of
this peak amplitude, $\max(|F_{\omega\neq0}|)$ on $H_{\mathrm{ext}}$ that
confirms the expectations.